\documentclass[12pt]{article}
\usepackage{graphicx}
\begin{document}

\begin{center}
{\LARGE \bf Lorentz group in ray optics}

\vspace{7mm}

Sibel Ba{\c s}kal \footnote{electronic
address:baskal@newton.physics.metu.edu.tr} \\
Department of Physics, Middle East Technical University,\\
06531 Ankara, Turkey
\vspace{5mm}

Elena Georgieva\footnote{electronic address:
egeorgie@pop500.gsfc.nasa.gov}\\
Science Systems and Applications, Inc., Lanham, MD 20771, U.S.A., and \\
National Aeronautics and Space Administration, \\Goddard Space
Flight Center, Laser and Electro-Optics Branch,\\ Code 554, Greenbelt,
Maryland 20771, U.S.A.

\vspace{5mm}

Y. S. Kim\footnote{electronic address: yskim@physics.umd.edu}\\
Department of Physics, University of Maryland,\\
College Park, Maryland 20742, U.S.A.\\

\vspace{5mm}

Marilyn E. Noz \footnote{electronic address: noz@nucmed.med.nyu.edu}\\
Department of Radiology, New York University, New York,
New York 10016, U.S.A.

\end{center}

\begin{abstract}
It has been almost one hundred years since Einstein formulated his
special theory of relativity in 1905.  He showed that the basic
space-time symmetry is dictated by the Lorentz group.  It is shown
that this group of Lorentz transformations is not only applicable to
special relativity, but also constitutes the scientific language for
optical sciences.   It is noted that coherent and squeezed states of
light are representations of the Lorentz group.  The Lorentz group is
also the basic underlying language for classical ray optics, including
polarization optics, interferometers, the Poincare\'e sphere, one-lens
optics, multi-lens optics, laser cavities, as well multilayer optics.
\end{abstract}

\newpage
\section{Introduction}\label{intro}

Since Einstein formulated special relativity in 1905, the basic
space-time symmetry has been that of the Lorentz group.  He established
the energy-momentum relation which is valid for slow massive particles,
high-speed massive particles, and massless particles.  Einstein
formulated this relation initially for particles without internal
space-time structures, but it is widely accepted that it is valid for
all particles including particles with spin and/or internal space-time
extension.

In 1939, Wigner published his most fundamental paper dealing with
internal space-time symmetries of relativistic particles~\cite{wig39}.
In this paper, Wigner introduced the Lorentz group to physics.
Furthermore, by introducing his ``little groups,'' Wigner provided the
framework for studying the internal space-time symmetries of relativistic
particles.  The scientific contents of this paper have not yet been
fully recognized by the physics community.  We are writing this report
as a continuation of the work Wigner initiated in this history-making
paper.

While particle physicists are still struggling to understand internal
space-time symmetries of elementary particles, Wigner's Lorentz group
is becoming useful to many other branches of physics.  Among them are
optical sciences, both quantum and classical.  In quantum optics, the
coherent and squeezed states are representations of the Lorentz
group~\cite{knp91}.  Recently, the Lorentz group is becoming the
fundamental language for classical ray optics.  It is gratifying to note
that optical components, such as lenses, polarizers, interferometers,
lasers, and multi-layers can all be formulated in terms of the Lorentz
group which Wigner formulated in his 1939 paper.  Classical ray optics
is of course a very old subject, but we cannot do new physics without
measurements using optical instruments.  Indeed, classical ray optics
constitutes Wigner's frontier in physics.

The word "group theory" sounds like abstract mathematics, but it is
gratifying to note that Wigner's little groups can be formulated in
terms of two-by-two matrices, while classical ray optics is largely a
physics of two-by-two matrices.  The mathematical correspondence is
straight-forward.

In order to see this point clearly, let us start with the
following classic example. The second-order differential equation
\begin{equation}
A {d^{2} q(t) \over dt^{2}} + B {d q(t) \over dt} +
C q(t) = F \cos(\omega t) .
\end{equation}
is applicable to a driven harmonic oscillator
with dissipation.  This can also be used for studying an electronic
circuits consisting of inductance, resistance, capacitance, and an
alternator.   Thus, it is possible to study the oscillator system using
an electronic circuit.  Likewise, an algebra of two-by-two matrices
can serve as the scientific language for several different branches of
physics, including special relativity, ray optics, and quantum optics.

\begin{table}
\caption{Further contents of Einstein's $E = mc^{2}$.  Massive
and massless particles have different energy-momentum relations.
Einstein's special relativity gives one relation for both.  Wigner's
little group unifies the internal space-time symmetries for massive and
massless particles.  The quark model and the parton model can also be
combined into one covariant model.}\label{einwig}
\begin{center}
\vspace{3mm}
\begin{tabular}{rccc}
\hline \\[-3.9mm]
\hline
{} & {} & {} & {}\\
{} & Massive, Slow \hspace*{1mm} & COVARIANCE \hspace*{1mm}&
Massless, Fast \\[4mm]\hline
{} & {} & {} & {}\\
Energy- & {}  & Einstein's & {} \\
Momentum & $E = p^{2}/2m_{0}$
 & $ E = \sqrt{p^{2}c^{2} + m_{0}^{2}c^{4}}$ & $E = cp$
\\[4mm]\hline
{} & {} & {} & {}\\
Internal & $S_{3} $ & {}  &  $S_{3}$ \\[-1mm]
space-time & {} & Wigner's  & {} \\ [-1mm]
symmetry & $S_{1}, S_{2}$ & Little Group & Gauge
Trans. \\[4mm]\hline
{} & {} & {} & {}\\
Relativistic & {} & Lorentz-squeezed  & {} \\[-1mm]
Extended &Quarks  & Harmonic    & Partons \\ [-1mm]
Particles & {} & Oscillators & {} \\[2mm]
\hline
\hline
\end{tabular}

\end{center}
\end{table}

There are many physical systems which can be formulated in terms of
two-by-two matrices.  If we restrict that their determinant be one,
there is a well established mathematical discipline called the group
theory of $SL(2,C)$. This aspect was noted in the study of Lorentz
transformations.  In group theoretical terminology, the group $SL(2,C)$
is the universal covering group of the group of the Lorentz group.
In practical terms, to each two-by-two matrix, there corresponds one
four-by-four matrix which performs a Lorentz transformation in the
four-dimensional Minkowskian space.  Thus, if a physical system can
be explained in terms of two-by-two matrices, it can be explained
with the language of Lorentz transformations.  Furthermore, the
physical system based on two-by-two matrices can serve as an analogue
computer for Lorentz transformations.

Optical filters, polarizers, and interferometers deal with two
independent optical rays.  They superpose the beams, change the
relative phase shift, and change relative amplitudes.  The basic
language here is called the Jones matrix formalism, consisting of
the two-by-two matrix representation of the $SL(2,C)$
group~\cite{hkn97,hkn99}.  The four-by-four Mueller matrices are
derivable from the two-by-two matrices of $SL(2,C)$.

For these two-beam systems, the Poincar\'e sphere serves as an
effective language~\cite{born80,bross98}.  Since the two-beam system
is described by the Lorentz group, the Poincar\'e sphere is
necessarily a representation of the Lorentz group.  We shall use
this sphere to study the degree of coherence between the beams.

Para-axial lens optics can also be formulated in terms of two-by-two
matrices, applicable to the two-component vector space consisting of
the distance from the optical axis and the slope with respect to the
axis.  The lens and translation matrices are triangular, but they
are basically representations of the $Sp(2)$ group which is the real
subgroup of the group $SL(2,C)$~\cite{bk01,bk03}.

Laser optics is basically multi-lens lens optics.  However, the problem
here is how to get a simple mathematical expression for the system of a
large number of the same lens separateled  by equal distance.  Here
again, group theory simplifies calculations~\cite{bk02}.

In multi-layer optics, we deal with two optical rays moving in opposite
directions.  The standard language in this case is the S-matrix
formalism~\cite{azzam77}. This is also a two-by-two matrix formalism.
As in the case of laser cavities, the problem is the multiplication of
a large number of matrix chains~\cite{gk01,gk03}.

It is shown in this report that the two-by-two representation of the
six-parameter Lorentz group is the underlying common scientific language
for all of the instruments mentioned above.  While the abstract group
theoretical ideas make two-by-two matrix calculations more systematic
and transparent in optics, optical instruments can act as analogue
computers for Lorentz transformations in special relativity.  It is
gratifying to note that special relativity and ray optics can be
formulated as the physics of two-by-two matrices.

In Sec.~\ref{lorentz}, we explain how the Lorentz group can be
formulated in terms of four-by-four matrices.  It is shown that the
group can have six independent parameters.  In Sec.~\ref{spinor}, we
explain how it is possible to formulate the Lorentz group in terms of
two-by-two matrices.  It is shown that the four-by-four transformation
matrices can be constructed from those those of the two-by-two
matrices.

In Sec.~\ref{wlittle}, we discuss the historical significance of
Wigner's 1939 paper~\cite{wig39} on the Lorentz group and its
application to the internal space-time symmetries of relativistic
particles. In Sec.~\ref{polari}, we present the basic building blocks
for the two-by-two representation of the Lorentz group in terms of the
matrices commonly seen in ray optics.

In Secs.~\ref{polari} and \ref{interfer}, we study the polarizations
for the two-beam system.  It is shown that both the Jones and Mueller
matrices are representations of the Lorentz group.  The role of the
Stokes parameters is discussed in detail.  It is shown in
Sec.~\ref{poinc} that the Poincar\'e sphere is a graphical
representation of the Poincar\'e group, and serves as a device to
describe the degree of coherence.

In Secs.~\ref{olens}, \ref{mlens}, \ref{lcav}, and \ref{mlayer}, we
discuss one-lens system, multi-lens system, laser cavities, and
multi-layer optics, respectively.  In all of these sections, the
central scientific language is the Lorentz group.

In Appendix~\ref{sqosc}, we expand the content of the third row of
Table~\ref{einwig}.  It is noted that the covariant harmonic oscillator
formalism can unify the quark model for slow hadrons with the parton
model for ultra-relativistic quarks.  It is then shown that the same
oscillator formalism serves as the basic scientific language for
squeezed states of light.

In Appendix~\ref{euler}, it is shown that the Lie-group method, in
terms of the generators, is not the only method in constructing group
representations.  For  the rotation group and the three-parameter
subgroups of the Lorentz group, it is simpler to start with the minimum
number of starter matrices.  For instance, while there are three
generators for the rotation group in the Lie approach, we can construct
the most general form of the rotation matrix from rotations around two
directions, as Goldstein constructed the Euler angles~\cite{gold80}.

In Appendix~\ref{conju}, it is noted that the four-by-four matrices are
real, their two-by-two counterparts are complex.  However, there is
a three-parameter real subgroup called $Sp(2)$.  It is shown that the
complex subgroup $SU(1,1)$ is equivalent to $Sp(2)$ through conjugate
transformation.

\section{Lorentz Transformations}\label{lorentz}
Let us consider the space-time coordinates $(ct, z, x, y)$.  Then the
rotation around the $z$ axis is performed by the four-by-four matrix
\begin{equation}\label{rotzz}
\pmatrix{1 & 0 & 0 & 0 \cr 0 & 0 & 0 & 0 \cr
0 & 0 & \cos\theta & -\sin\theta \cr 0 & 0 & \sin\theta & \cos\theta} .
\end{equation}
This transformation is generated by
\begin{equation}\label{j33}
J_{3} = \pmatrix{0 & 0 & 0 & 0 \cr 0 & 0 & 0 & 0 \cr
0 & 0 & 0 & -i \cr 0 & 0 & i & 0} .
\end{equation}
Likewise, we can write down the generators of rotations $J_{1}$ and
$J_{2}$ around the $x$ and $y$ axes respectively.
\begin{equation}\label{j12}
J_{1} = \pmatrix{0 & 0 & 0 & 0 \cr 0 & 0 & 0 & i \cr
0 & 0 & 0 & 0 \cr 0 & -i & 0 & 0} , \qquad
J_{2} = \pmatrix{0 & 0 & 0 & 0 \cr 0 & 0 & -i & 0 \cr
0 & i & 0 & 0 \cr 0 & 0 & 0 & 0} .
\end{equation}
These three generators
satisfy the closed set of commutations relations
\begin{equation}\label{rotcom}
\left[J_{i}, J_{j} \right] = i \epsilon_{ijk} J_{k} .
\end{equation}
This set of commutation relations is for the three-dimensional
rotation group.

The Lorentz boost along the $z$ axis takes the form
\begin{equation}\label{boostz}
B = \pmatrix{\cosh\eta & \sinh\eta & 0 & 0 \cr
\sinh\eta & \cosh\eta & 0 & 0 \cr 0 & 0 & 1 & 0 \cr 0 & 0 & 0 & 1} ,
\end{equation}
which is generated by
\begin{equation}\label{k33}
K_{3} = \pmatrix{0 & i & 0 & 0 \cr i & 0 & 0 & 0 \cr
0 & 0 & 0 & 0 \cr 0 & 0 & 0 & 0} .
\end{equation}
Likewise, we can write generators of boosts $K_{1}$ and $K_{2}$
along the $x$ and $y$ axes respectively, and they take the form
\begin{equation}\label{k12}
K_{1} = \pmatrix{0 & 0 & i & 0 \cr 0 & 0 & 0 & 0 \cr
i & 0 & 0 & 0 \cr 0 & 0 & 0 & 0} , \qquad
K_{2} = \pmatrix{0 & 0 & 0  & i \cr 0 & 0 & 0 & 0 \cr
0 & 0 & 0 & 0 \cr i & 0 & 0 & 0} .
\end{equation}
These boost generators satisfy the commutation relations
\begin{equation}\label{locom}
\left[J_{i}, K_{j} \right] = i \epsilon_{ijk} K_{k} , \qquad
\left[K_{i}, K_{j} \right] = -i \epsilon_{ijk} J_{k} .
\end{equation}

Indeed, the three rotation generators and the three boost generators
satisfy the closed set of commutation relations given in
Eq.(\ref{rotcom}) and Eq.(\ref{locom}).  These three commutation
relations form the starting point of the Lorentz group.  The generators
given in this section are four-by-four matrices, but they are not the
only set satisfying the commutation relations.   We can construct also
six two-by-two matrices satisfying the same set of commutation relations.
The group of transformations constructed from these two-by-two matrices
is often called $SL(2,c)$ or the two-dimensional representation of the
Lorentz group.  Throughout the present paper, we used the two-by-two
transformation matrices constructed from the generators of the
$SL(2,c)$ group.

\section{Spinors and Four-vectors}\label{spinor}
In Sec.~\ref{lorentz}, we started with four-by-four transformation
matrices applicable to the four-dimensional space-time.  We then ended
up with a set of closed commutation relations for the six generators
consisting of three rotation and three boost generators.  These
generators are in the form of four-by-four matrices.

In this section, we shall see first that there is a set of two-by-two
matrices satisfying the same set of commutation relations, constituting
the two-by-two representations of the Lorentz group.  The representation
so constructed is called the $SL(2,c)$ group, or the universal covering
group of the Lorentz group.  The transformation matrices are applicable
to two-component $SL(2,C)$ spinors.  The algebraic property of this
two-by-two representation is the same as that of the four-by-four
representation.

As we shall see in this paper, the spinors and the four-vectors
correspond to Jones vectors and Stokes parameters respectively in
polarization optics~\cite{hkn99}.  The question then is whether we can
construct the four-vector from the spinors.  In the language of
polarization optics, the question is whether it is possible to
construct the coherency matrix~\cite{born80,bross98} from the Jones
vector.

With this point in mind, let us start from the following form of the
Pauli spin matrices:
\begin{equation}
\sigma_{1} = \pmatrix{1 & 0 \cr 0 & -1} , \quad
\sigma_{2} = \pmatrix{0 & 1 \cr 1 & 0} , \quad
\sigma_{3} = \pmatrix{0 & -i \cr i & 0} .
\end{equation}
These matrices are written in a different convention.  Here
$\sigma_{3}$ is imaginary, while $\sigma_{2}$ is imaginary in the
traditional notation.  Also in this convention, we can construct
three rotation generators
\begin{equation}
J_{i} = {1 \over 2} \sigma_{i} ,
\end{equation}
which satisfy the closed set of commutation relations
\begin{equation}\label{comm1}
\left[J_{i}, J_{j}\right] = i \epsilon_{ijk} J_{k} .
\end{equation}
We can also construct three boost generators
\begin{equation}
K_{i} = {i \over 2} \sigma_{i} ,
\end{equation}
which satisfy the commutation relations
\begin{equation}\label{comm2}
\left[K_{i}, K_{j}\right] = -i \epsilon_{ijk} J_{k} .
\end{equation}
The $K_{i}$ matrices alone do not form a closed set of commutation
relations, and the rotation generators $J_{i}$ are needed to form a
closed set:
\begin{equation}\label{comm3}
\left[J_{i}, K_{j}\right] = i \epsilon_{ijk} K_{k} .
\end{equation}

The six matrices $J_{i}$ and $K_{i}$ form a closed set of commutation
relations, and they are like the generators of the Lorentz group
applicable to the (3 + 1)-dimensional Minkowski space.  The group
generated by the above six matrices is called $SL(2,c)$ consisting of
all two-by-two complex matrices with unit determinant.

Let us write the two-by-two transformation matrix as
\begin{equation}\label{alpha}
L = \pmatrix{\alpha &\beta \cr \gamma  & \delta} .
\end{equation}
This matrix has four complex elements with eight real parameters.
However, the six generators are all traceless and, the determinant of
the matrix has to be one.  Thus, it has six independent parameters.

In order to construct four-vectors, we need two different spinor
representations of the Lorentz group.  Let us go to the commutation
relations for the generators given in Eqs.(\ref{comm1}), (\ref{comm2})
and (\ref{comm3}).  These commutators are not invariant under the sign
change of the rotation generators $J_{i}$, but are invariant under the
sign change of the squeeze operators $K_{i}$.  Thus, to each spinor
representation, there is another representation with the squeeze
generators with opposite sign.  This allows us to construct another
representation with the generators:
\begin{equation}
\dot{J}_{i} = {1 \over 2} \sigma_{i}, \qquad
\dot{K}_{i} = {-i \over 2} \sigma_{i} .
\end{equation}
We call this representation the ``dotted'' representation.  If we write
the transformation matrix $L$ of Eq.(\ref{alpha}) in terms of the
generators as
\begin{equation}
L = \exp\left\{-{i\over 2} \sum_{i=1}^{3}\left(\theta_{i}\sigma_{i} +
i\eta_{i}\sigma_{i}\right) \right\} ,
\end{equation}
then the transformation matrix in the dotted representation becomes
\begin{equation}\label{eldot}
\dot{L} = \exp\left\{-{i\over 2} \sum_{i=1}^{3}\left(\theta_{i}\sigma_{i}
- i\eta_{i}\sigma_{i}\right)\right\} .
\end{equation}
In both of the above matrices, Hermitian conjugation changes the
direction of rotation.  However, it does not change the direction of
boosts.  We can achieve this only by changing $L$ to $\dot{L}$,
and we shall call this the ``dot'' conjugation.

Likewise, there are two different set of spinors.  Let us use write
\begin{equation}\label{uv}
\pmatrix{u \cr v}, \quad \pmatrix{\dot{u}  \cr \dot{v} }
\end{equation}
for the spinor in the undotted and dotted representations respectively.
Then  the four-vectors are constructed as~\cite{hks86,hks86jm}
\begin{eqnarray}
&{}&  u\dot{u} = - (x - iy), \quad v\dot{v} = (x + iy), \nonumber \\[2ex]
&{}&   u\dot{v} = (t + z), \quad v\dot{u} = -(t - z)
\end{eqnarray}
leading to the matrix
\begin{equation}\label{dotmat}
V = \pmatrix{u \dot{v} & -u\dot{u} \cr v\dot{v} & -v\dot{u}}
   = \pmatrix{u \cr v} \pmatrix {\dot{v} & -\dot{u}} ,
\end{equation}
where $u$ and $\dot{u}$ are one if the spin is up, and are zero if the
spin is down, while $v$ and $\dot{v}$ are zero and one for the spin-up
and spin-down cases.

If the two-by-two matrix of Eq.~\ref{alpha} is applicable to the column
vector of Eq.(\ref{uv}), what is the transformation matrix applicable
to the row vector $(\dot{v},~-\dot{u})$ from the right-hand side?  It is
the transpose of the matrix applicable to the column vector
$(\dot{v},~-\dot{u})$.  We can obtain this column vector from
\begin{equation}\label{dotcol}
 \pmatrix {\dot{v} \cr -\dot{u}} ,
\end{equation}
by applying to it the matrix
\begin{equation}
g = -i\sigma_{3} = \pmatrix{0 & -1 \cr 1 & 0} .
\end{equation}
This matrix also has the property
\begin{equation}
g \sigma_{i} g^{-1} = -\left(\sigma_{i}\right)^{T} ,
\end{equation}
where the superscript $T$ means the transpose of the matrix.  The
transformation matrix applicable to the column vector of
Eq.(\ref{dotcol}) is $\dot{L}$ of Eq.(\ref{eldot}).  Thus the matrix
applicable to the row vector $(\dot{v},~-\dot{u})$ in Eq.(\ref{dotmat})
is
\begin{equation}
\left\{g^{-1} L g\right\}^{T} = g^{-1} L^{T} g .
\end{equation}
This is precisely the Hermitian conjugate of $L$.

Let us now write the $V$ matrix of Eq.(\ref{dotmat}) as
\begin{equation}\label{matV}
V = \pmatrix{t + z & x - iy \cr x + iy & t - z} ,
\end{equation}
where the set of variables $(x, y, z, t)$ is transformed like a
four-vector under Lorentz transformations.  Then the Lorentz
transformation on $V$ can be performed as
\begin{equation}\label{Ldag}
V' = L V L^{\dagger} ,
\end{equation}
where the transformation matrix $L$ is that of Eq.(\ref{alpha}).
As we have seen in this section, the construction of four-vectors
from the two-component spinors is not a trivial task, but has been
discussed in the literature~\cite{hks86,bk95}.  Likewise, it is
possible to construct four-by-four Lorentz transformation matrices
from the two-by-two matrix of Eq.(\ref{alpha})~\cite{bk95,knp86}.

As we shall see in the present paper, the matrix of the form
Eq.(\ref{dotmat}) or Eq.(\ref{matV}) appears in optics as the coherency
matrix and the density matrix for two beam systems such as polarization
optics and interferometers.

\section{Wigner's Little Groups}\label{wlittle}
In his 1939 paper, Wigner introduced his little groups in order to
study the internal space-time structures of relativistic particles.
The little group is the maximal subgroup of the Lorentz group which
leaves the momentum of the particle invariant.  The little groups,
originally developed for studying relativistic symmetries of particles,
are now becoming an important scientific language for ray optics as
we shall see in this paper.

Let us first discuss the physical basis of Wigner's little groups.
If the speed of a particle is much smaller than that of light,
energy-momentum relation is $E = p^{2}/2m_{0} + m_{0}c^{2}$.  If the
speed is close to that of light, the relation is $E = cp$.  These two
different relations can be combined into one covariant formula
$E^{2} = m_{0}^{2}c^{4} + p^{2}c^{2}$.  This aspect of Einstein's
$E = mc^{2}$ is well known, as indicated in Table~\ref{einwig}.

In addition, particles have internal space-time variables.  Massive
particles have spins while massless particles have their helicities
and gauge variables.  Our first question is whether this aspect of
space-time variables can be unified into one covariant concept.
The answer to this question is Yes.  Wigner's little group does the
job, also as indicated Table~\ref{einwig}.

particles can also have space-time extensions.  For instance, in the
quark model, hadrons are bound states of quarks.  However, the hadron
appears as a collection of partons when it moves with speed close to
the velocity of light.  Quarks and partons seem to have quite distinct
properties.  Are they different manifestations of a single covariant
entity?  This is one of the most pressing issues in high-energy
particles physics.  The third row of Table~\ref{einwig} addresses this
question.

The mathematical framework of this program was developed by Eugene
Wigner in 1939~\cite{wig39}.  He constructed the maximal subgroups of
the Lorentz group whose transformations will leave the four-momentum
of a given particle invariant.  These groups are known as Wigner's
little groups.  Thus, the transformations of the little groups change
the internal space-time variables of the particle, while leaving its
momentum invariant.

In his paper~\cite{wig39}, Wigner shows that, for each massive
particle, there is a Lorentz frame in which the particle is at rest.
Then the three-dimensional rotation group leaves its momentum invariant,
while changing the direction of its spin.  Thus, the little group for
this particular case is $O(3)$.  In other Lorentz frames, the little
group is a Lorentz-boosted $O(3)$ group.

For a massless particle, Wigner notes that there are no Lorentz frames
in which the particle is at rest.  The best we can do is to align its
momentum to a given axis, or the $z$ axis.  Then, its momentum is
invariant under rotations around the $z$ axis.  In addition, Wigner
found two more transformations which leave the momentum invariant.
The physics of these additional degrees was not explained in Wigner's
original paper~\cite{wig39}, but their mathematics has been worked out.
The physics of these degrees of freedom was later determined to
be that of gauge transformations~\cite{janner71}, and its complete
understanding was not achieved until 1990~\cite{kiwi90jm}.

The group of Lorentz transformations consists of three boosts and
three rotations.  The rotations therefore constitute a subgroup of
the Lorentz group.  If a massive particle is at rest, its four-momentum
is invariant under rotations.  Thus the little group for a massive
particle at rest is the three-dimensional rotation group.  Then what is
affected by the rotation?  The answer to this question is very simple.
The particle in general has its spin.  The spin orientation is going
to be affected by the rotation!  If we use the four-vector coordinate
$(ct, z, x, y)$, the four-momentum vector for the particle at rest is
$(m_{0}c ,0 , 0, 0)$, and the three-dimensional rotation group leaves
this four-momentum invariant.  This little group is generated by
the three rotation generators given in Eq.(\ref{j33}) and Eq.(\ref{j12}).
They satisfy the commutation relations for the rotation group given
Eq.(\ref{rotcom}).

If the rest-particle is boosted along the $z$ direction, it
up a non-zero momentum component along the same direction.  The $O(3)$
generators will also be boosted.  The boost matrix takes the form of
Eq.(\ref{boostz}), and the boosted generators will be
\begin{equation}\label{boost}
J_{i}' = B~J_{i}~B^{-1} ,
\end{equation}
and this boost will not change the commutation relation of
Eq.(\ref{rotcom}).

For a massless particle moving along the $Z$ direction, Wigner observed
that the little group generated by the rotation generator around the
$z$ axis, namely $J_{3}$ of Eq.(\ref{j33}), and two other generators
which take the form
\begin{equation}\label{n1n2}
N_{1} = \pmatrix{0 & 0 & i & 0 \cr 0 & 0 & i & 0
\cr i & -i & 0 & 0 \cr 0 & 0 & 0 & 0} ,  \qquad
N_{2} = \pmatrix{0 & 0 & 0 & i \cr 0 & 0 & 0 & i
\cr 0 & 0 & 0 & 0 \cr i & -i & 0 & 0} .
\end{equation}
If we use $K_{i}$ for the boost generator along the i-th axis, these
matrices can be written as
\begin{equation}
N_{1} = K_{1} - J_{2} , \qquad N_{2} = K_{2} + J_{1} ,
\end{equation}
with $K_{1}$ and $K_{2}$ given in Eq.(\ref{k12}).

The generators $J_{3}, N_{1}$ and $N_{2}$ satisfy the following set
of commutation relations.
\begin{equation}\label{e2lcom}
[N_{1}, N_{2}] = 0 , \qquad [J_{3}, N_{1}] = iN_{2} ,
\qquad [J_{3}, N_{2}] = -iN_{1} .
\end{equation}

In order to understand the mathematical basis of the above commutation
relations, let us consider transformations on a two-dimensional plane
with the $xy$ coordinate system.  We can then make rotations around
the origin and translations along the $x$ and $y$ directions.  If we
write these generators as $L, P_{x}$ and $P_{y}$ respectively, they
satisfy the commutation relations~\cite{knp86}
\begin{equation}\label{e2com}
[P_{x}, P_{y}] = 0 , \qquad [L, P_{x}] = iP_{y} ,
\qquad [L, P_{y}] = -iP_{x} .
\end{equation}
This is a closed set of commutation relations for the generators of the
$E(2)$ group, or the two-dimensional Euclidean group.  If we replace
$N_{1}$ and $N_{2}$ of Eq.(\ref{e2lcom}) by $P_{x}$ and $P_{y}$, and
$J_{3}$ by $L$, the commutations relations for the generators of the
$E(2)$-like little group becomes those for the $E(2)$-like little group.
This is precisely why we say that the little group for massless
particles are like $E(2)$.

It is not difficult to associate the rotation generator $J_{3}$ with
the helicity degree of freedom of the massless particle.   Then what
physical variable is associated with the $N_{1}$ and $N_{2}$
generators?  Indeed, Wigner was the one who discovered the existence
of these generators, but did not give any physical interpretation to
these translation-like generators in his original paper~\cite{wig39}.
For this reason, for many years, only those representations with the
zero-eigenvalues of the $N$ operators were thought to be physically
meaningful representations~\cite{wein64}.  It was not until 1971
when Janner and Janssen reported that the transformations generated
by these operators are gauge transformations~\cite{janner71}.  The
role of this translation-like transformation has also been studied
for spin-1/2 particles, and it was concluded that the polarization
of neutrinos is due to gauge invariance~\cite{hks82}.

The $O(3)$-like little group remains $O(3)$-like when the particle is
Lorentz-boosted.  Then, what happens when the particle speed becomes
the speed of light?  The energy-momentum relation
$E^{2} = m_{0}^{2}c^{4} + p^{2}c^{2}$ become $E = pc$.
Is there then a limiting case of the $O(3)$-like little group?  Since
those little groups are like the three-dimensional rotation group
and the two-dimensional Euclidean group respectively, we are first
interested in whether $E(2)$ can be obtained from $O(3)$.  This will
then give a clue as to how to obtain the $E(2)$-like little group as
a limiting case of $O(3)$-like little group.  With this point in mind,
let us look into this geometrical problem.

In 1953, In{\"o}n{\"u} and Wigner formulated this problem as the
contraction of $O(3)$ to $E(2)$~\cite{inonu53}.  Let us see what they
did.  We always associate the three-dimensional rotation group with a
spherical surface.  Let us consider a circular area of radius one
kilometer centered on the north pole of the earth.  Since the radius
of the earth is more than 6,450 times longer, the circular region
appears to be flat.  Thus, within this region, we use the $E(2)$
symmetry group.  The validity of this approximation depends on the
ratio of the two radii.

How about then the little groups which are isomorphic to $O(3)$ and
$E(2)$?  It is reasonable to expect that the $E(2)$-like little group
can be obtained as a limiting case of the $O(3)$-like little group
for massless particles.  In 1981, it was observed by Bacry and
Chang~\cite{bacry68} and by Ferrara and Savoy~\cite{ferrara82} that
this limiting process is the Lorentz boost to infinite-momentum frame.

In 1983, it was noted by Han {\it et al} that the large-radius limit
in the contraction of $O(3)$ to $E(2)$ corresponds to the
infinite-momentum limit for the case of the $O(3)$-like little group
to the $E(2)$-like little group for massless particles.  They showed
that transverse rotation generators become the generators of gauge
transformations in the limit of infinite momentum~\cite{hks83pl}.

Let us see how this happens. The $J_{3}$ operator of Eq.(\ref{j33}),
which generates rotations around the $z$ axis, is not affected by
the boost conjugation of the $B$ matrix of Eq.(\ref{boost}).  On
the other hand, the $J_{1}$ and $J_{2}$ matrices become
\begin{equation}
N_{1} = \lim_{\eta\rightarrow \infty}e^{-\eta} B^{-1} J_{2} B ,
\qquad
N_{2} = \lim_{\eta\rightarrow \infty}-e^{-\eta} B^{-1} J_{1} B ,
\end{equation}
which are given in Eq.(\ref{n1n2}).  The generators $N_{1}$ and
$N_{2}$ are the contracted $J_{2}$ and $J_{1}$ respectively in the
infinite-momentum.  In 1987, Kim and Wigner studied this problem in
more detail and showed that the little group for massless particles
is the cylindrical group which is isomorphic to the $E(2)$
group~\cite{kiwi87jm}.

This completes the second row in Table~\ref{einwig}, where Wigner's
little group unifies the internal space-time symmetries of massive
and massless particles.  The transverse components of the rotation
generators become generators of gauge transformations in the
infinite-momentum limit.

Let us go back to Table I given in Sec.~\ref{intro}.  As for the third
row for relativistic extended particles, the most efficient approach
is to construct representations of the little groups using the wave
functions which can be Lorentz-boosted.  This means that we have to
construct wave functions which are consistent with all known rules of
quantum mechanics and special relativity.  It is possible to construct
harmonic oscillator wave functions which satisfy these conditions.
We can then take the low-speed and high-speed limits of the covariant
harmonic oscillator wave functions for the quark model and the parton
model  respectively.  This aspect was extensively discussed in the
literature~\cite{knp86,kn73}, and is beyond the scope of the present
report.

However, it is important to note that the covariant harmonic oscillator
formalism use for this purpose can serve as the Fock space description
for the squeezed state of light~\cite{yuen76}, we give a brief
discussion of this aspect in Appendix~\ref{sqosc}, entitled ``Squeezed
Harmonic Oscillators.''

In this section, we discussed Wigner's little groups applicable to
internal space-time symmetries of relativistic particles.  However,
as we shall see in this paper, these little groups play an important
role in understanding ray optics.  Conversely, optical configurations
in ray optics can serve as analogue computers for space-time symmetries
in particle physics.

\section{Polarization Optics}\label{polari}
Let us consider two optical beams propagating along the $z$ axis.  We
are then led to the column vector:
\begin{equation}\label{jones}
\pmatrix{\psi_{1}(z, t) \cr \psi_{2}(z, t) } =
\pmatrix{A_{1}~\exp{\left(-i\left(kz -\omega t + \phi_{1}\right)\right) }
\cr A_{2}~\exp{\left(-i\left(kz -\omega t + \phi_{2}\right)\right) } } .
\end{equation}
We can then achieve a phase shift between the beams by applying the
two-by-two matrix:
\begin{equation}\label{rot11}
\pmatrix{e^{i\phi/2} & 0 \cr 0 & e^{-i\phi/2}}.
\end{equation}
If we are interested in mixing up the two beams, we can apply
\begin{equation}\label{rot22}
\pmatrix{\cos(\theta/2)  & -\sin(\theta/2)  \cr
\sin(\theta/2)  & \cos(\theta/2)}
\end{equation}
to the column vector.

If the amplitudes become changed either by attenuation or by reflection,
we can use the matrix
\begin{equation}\label{boost11}
\pmatrix{e^{\eta/2}  & 0  \cr 0 & e^{-\eta/2}}
\end{equation}
for the change.  In this paper, we are dealing only with the relative
amplitudes, or the ratio of the amplitudes.  As we shall see in
Sec.~\ref{interfer}, the above two-by-two matrices have their
corresponding four-by-four matrices respectively.

Repeated applications of these matrices lead to the form
\begin{equation}\label{alpha2}
G = \pmatrix{\alpha & \beta \cr \gamma  & \delta} ,
\end{equation}
where the elements are in general complex numbers.  The determinant
of this matrix is one.  Thus, the matrix can have six independent
parameters.

This matrix takes the identical form as the two-by-two matrix given in
Eq.(\ref{alpha2}).  However, the construction process is different.
The Lie-group generators are used for Eq.(\ref{alpha2}), while
Eq.(\ref{alpha2}) is constructed from repeated applications of the
transformation matrices.  The difference between these two methods
is discussed in Appendix~\ref{euler}.

In either case, this matrix is the most general form of the matrices
in the $SL(2,c)$ group, which is known to be the universal
covering group for the six-parameter Lorentz group.  This means
that, to each two-by-two matrix of $SL(2,c)$, there corresponds
one four-by-four matrix of the group of Lorentz transformations
applicable to the four-dimensional Minkowski space~\cite{knp86}.
It is possible to construct explicitly the four-by-four
Lorentz transformation matrix from the parameters $\alpha, \beta,
\gamma,$ and $\delta$.  This expression is available in the
literature~\cite{knp86}, and we consider here only special cases.

The four-by-four representation of the Lorentz group can be constructed
from the two-by-two representation of the $SL(2,c)$ group, which is
known as the universal covering group of the six-parameter Lorentz
group.  This aspect is discussed in Sec.~\ref{spinor}.

Let us go back to Eq.(\ref{alpha2}), the $SL(2,c)$ group represented by
this matrix has many interesting subgroups.  If the matrices are to be
Hermitian, then the subgroup is $SU(2)$ corresponding to the
three-dimensional rotation group.  If all the elements are real numbers,
the group becomes the three-parameter $Sp(2)$ group.  This subgroup is
equivalent to $SU(1,1)$ which is the primary scientific language for
squeezed states of light~\cite{knp91,yuen76}.

We can also consider the matrix of Eq.(\ref{alpha2}) when one of its
off-diagonal elements vanishes.  Then, it takes the form
\begin{equation}
   \pmatrix{\exp{(i\phi/2)}   &    0  \cr \gamma &
   \exp{(-i\phi/2)} } ,
\end{equation}
where $\gamma$ is a complex number with two real parameters.  In
1939~\cite{wig39}, Wigner observed this form as one of the subgroups
of the Lorentz group.  He observed further that this group is isomorphic
to the two-dimensional Euclidean group, and that its four-by-four
equivalent can explain the internal space-time symmetries of massless
particles including the photons.

In ray optics, we often have to deal with this type of triangular
matrices, particularly in lens optics and stability problems in laser
and multi-layer optics.  In the language of mathematics, dealing with
this form is called the Iwasawa decomposition~\cite{hkn99,iwa49}.

If the Jones matrix contains all the parameters for the polarized
light beam, why do we need the mathematics in the four-dimensional
space? The answer to this question is well known.  In addition to the
basic parameter given by the Jones vector, the Stokes parameters give
the degree of coherence between the two rays.

Let us go back to the Jones spinor of Eq.(\ref{jones}), and construct
the quantities:
\begin{eqnarray}
&{}&  S_{11} = <\psi_{1}^{*}\psi_{1}> , \qquad
  S_{12} = <\psi_{1}^{*}\psi_{2}> , \nonumber \\[2ex]
&{}&  S_{21} = <\psi_{2}^{*}\psi_{1}> , \qquad
  S_{22} = <\psi_{2}^{*}\psi_{2}>  .
\end{eqnarray}
Then the Stokes vector consists of
\begin{eqnarray}
&{}&  S_{0} = S_{11} + S_{22} , \quad
  S_{1} = S_{11} - S_{22}  , \nonumber \\[2ex]
&{}&  S_{2} = S_{12} + S_{21}  , \quad
  S_{3} = -i(S_{11} + S_{22})   .
\end{eqnarray}
The four-component vector
\begin{equation}
(S_{0}, S_{1}, S_{2}, S_{3})
\end{equation}
transforms like the space-time four-vector $(ct, z, x, y)$ under
Lorentz transformations.  The Mueller matrix is therefore like the
Lorentz-transformation matrix.

As in the case of special relativity, let us consider the quantity
\begin{equation}
M^{2} = S_{0}^{2}- S_{1}^{2} - S_{2}^{2} - S_{3}^{2} .
\end{equation}
Then $M$ is like the mass of the particle while the Stokes four-vector
is like the four-momentum.

If $M = 0$,  the two-beams are in a pure state. As $M$ increases,
the system becomes mixed, and the entropy increases. If it reaches the
value of $S_{0}$, the system becomes completely random.  It is
gratifying to note that this mechanism can be formulated in terms of
the four-momentum in particle physics~\cite{hkn99}.

Although we can borrow all the elegant mathematical identities of
the two-by-two representations of the Lorentz group, this formalism
does not allow us to describe the loss of coherence within the
interferometer system.  In order to study this effect, we have to
construct the coherency matrix:
\begin{equation}\label{cocy}
C = \pmatrix{S_{11} & S_{12} \cr S_{21} & S_{22}}.
\end{equation}
Under the optical transformations discussed in this section, the
coherency matrix is transformed as
\begin{equation}\label{lt22}
C' = G~C~G^{\dagger},
\end{equation}
as in the case of the Lorentz transformation given in Eq.(\ref{Ldag}).

Using this formalisms based on the Lorentz group, we can discuss
the group theoretical property of polarization optics in
detail~\cite{yskim00}.  However, polarization optics is based on
two independent beams propagating in the same direction.  Since
interferometers are also based on the same optical system, we can
continue our discussion in the following section on interferometers.

\section{Interferometers}\label{interfer}
Typically, one beam is divided into two by a beam splitter.  We can
write the incoming beam as
\begin{equation}\label{expo1}
\Psi = \pmatrix{\psi_{1} \cr \psi_{2}} =
\pmatrix{ \exp{\left\{i(kz - \omega t)\right\}} \cr 0} .
\end{equation}
Then, the beam splitter can be written in the form of a rotation
matrix~\cite{sand99}:
\begin{equation}\label{rot222}
R(\theta) = \pmatrix{\cos(\theta/2) & -\sin(\theta/2) \cr
\sin(\theta/2) & \cos(\theta/2) } ,
\end{equation}
which transforms the column vector of Eq.(\ref{expo1}) into
\begin{equation}\label{expo2}
\pmatrix{\psi_{1} \cr \psi_{2}} =
\pmatrix{[\cos(\theta/2)]\exp{\left\{i(kz - \omega t)\right\}} \cr
-[\sin(\theta/2)]\exp{\left\{i(kz - \omega t)\right\}} } .
\end{equation}
The first beam $\psi_{1}$ of Eq.(\ref{expo1}) is now split into
$\psi_{1}$ and $\psi_{2}$ of Eq.(\ref{expo2}).  The intensity is
conserved.  If the rotation angle $\theta$ is -$\pi/4$, the
initial beam is divided into two beams of the same intensity and
the same phase~\cite{campos89}.

These two beams go through two different optical path lengths,
resulting in a phase difference.  If the phase difference is
$\phi$, the phase shift matrix is
\begin{equation}\label{shif22}
P(\phi) = \pmatrix{e^{-i\phi/2} & 0 \cr 0 & e^{i\phi/2}} .
\end{equation}

When reflected from mirrors, or while going through beam splitters,
there are intensity losses for both beams.  The rate of loss is not
the same for the beams.  This results in the attenuation matrix
of the form
\begin{equation}\label{atten}
\pmatrix{e^{-\eta_{1}} & 0 \cr 0 & e^{-\eta_{2}}} =
e^{-(\eta_{1} + \eta_{2})/2} \pmatrix{e^{\eta/2} & 0 \cr 0 &
e^{-\eta/2}}
\end{equation}
with $\eta = \eta_{2} - \eta_{1}$ .
This attenuator matrix tells us that the electric fields are
attenuated at two different rates.  The exponential factor
$e^{-(\eta_{1} + \eta_{2})/2}$ reduces both components at the same
rate and does not affect the degree of polarization.  The effect of
polarization is solely determined by the squeeze matrix
\begin{equation}\label{sq22}
S(\eta) = \pmatrix{e^{\eta/2} & 0 \cr 0 & e^{-\eta/2}} .
\end{equation}

In the detector or in the beam synthesizer, the two beams undergo a
superposition.  This can be achieved by the rotation matrix like the
one given in Eq.(\ref{rot22})~\cite{sand99}.   For instance, if the
angle $\theta$ is $90^{o}$, the rotation matrix takes the form
\begin{equation}
{1 \over \sqrt{2}}\pmatrix{1 & -1 \cr 1 & 1} .
\end{equation}
If this matrix is applied to the column vector of Eq.(\ref{expo2}),
the result is
\begin{equation}
{1 \over \sqrt{2}} \pmatrix{\psi_{1} - \psi_{2} \cr
\psi_{1} + \psi_{2}} .
\end{equation}
The upper and lower components show the interferences with negative
and positive signs respectively.

We have shown previously~\cite{hkn97} that the four-by-four
transformation matrices applicable to the Stokes parameters are
like Lorentz-transformation matrices applicable to the space-time
Minkowskian vector $(ct, z, x, y)$.  This allows us to study
space-time symmetries in terms of the Stokes parameters which are
applicable to interferometers.  Let us first see how the rotation
matrix of Eq.(\ref{rot22}) is translated into the four-by-four
formalism.  In this case,
\begin{equation}
\alpha = \delta = \cos(\theta/2), \qquad
\gamma = -\beta = \sin(\theta/2) .
\end{equation}
The corresponding four-by-four matrix takes the form~\cite{hkn99}
\begin{equation}\label{rot44}
R(\theta) = \pmatrix{1 & 0 & 0 & 0 \cr
0 & \cos\theta & -\sin\theta & 0  \cr
0 & \sin\theta & \cos\theta & 0 \cr
0 & 0 & 0 & 1} .
\end{equation}

Let us next see how the phase-shift matrix of Eq.(\ref{shif22}) is
translated into this four-dimensional space.  For this two-by-two
matrix,
\begin{equation}
\alpha = e^{-i\phi/2} , \qquad \beta = \gamma = 0 , \qquad
\delta = e^{i\phi/2} .
\end{equation}
For these values, the four-by-four transformation matrix
takes the form~\cite{hkn99}
\begin{equation}\label{shif44}
P(\phi) = \pmatrix{1 & 0 & 0 & 0 \cr 0 & 1 & 0 & 0  \cr
0 & 0 & \cos\phi & -\sin\phi \cr 0 & 0 & \sin\phi & \cos\phi} .
\end{equation}
For the squeeze matrix of Eq.(\ref{sq22}),
\begin{equation}
\alpha = e^{\eta/2}, \qquad \beta = \gamma = 0 , \qquad
\delta = e^{-\eta/2} .
\end{equation}
As a consequence, its four-by-four equivalent is
\begin{equation}\label{sq44}
S(\eta) = \pmatrix{\cosh\eta & \sinh\eta & 0 & 0 \cr
\sinh\eta & \cosh\eta & 0 & 0 \cr
0 & 0 & 1 & 0 \cr 0 & 0 & 0 & 1} .
\end{equation}
If the above matrices are applied to the four-dimensional Minkowskian
space of $(ct, z, x, y)$, the above squeeze matrix will perform a
Lorentz boost along the $z$ or $S_{1}$ axis with $S_{0}$ as the time
variable.  The rotation matrix of Eq.(\ref{rot44}) will perform a
rotation around the $y$ or $S_{3}$ axis, while the phase shifter of
Eq.(\ref{shif44}) performs a rotation around the $z$ or the $S_{1}$
axis.  Matrix multiplications with $R(\theta)$ and $P(\phi)$ lead to
the three-parameter group of rotation matrices applicable to the
three-dimensional space of $(S_{1}, S_{2}, S_{3})$.

The phase shifter  $P(\phi)$ of Eq.(\ref{shif44}) commutes with the
squeeze matrix of Eq.(\ref{sq44}), but the rotation matrix $R(\theta)$
does not.  This aspect of matrix algebra leads to many interesting
mathematical identities which can be tested in laboratories.  One of
the interesting cases is that we can produce a rotation by performing
three squeezes~\cite{hkn99}.  Another interesting case is a combination
of squeeze and rotation matrices which will lead to a triangular matrix
with unit diagonal elements.  This aspect is known as the Iwasawa
decomposition and is discussed in detail in Ref.~\cite{hkn99}.

\section{Density Matrices and Their Little Groups}\label{denma}
According to the definition of the density matrix~\cite{fey72}, the
coherency matrix of Eq.(\ref{cocy}) is also the density matrix.  In
this section, we shall discuss the cohrency matrix as the density
matrix.

Under the influence of the $G$ transformation given in Eq.(\ref{lt22}),
this coherency matrix is transformed as
\begin{eqnarray}\label{trans22}
&{}& C' = G\,C\,G^{\dagger} =
\pmatrix{S'_{11} & S'_{12} \cr S'_{21} & S'_{22}}  \nonumber \\[2ex]
&{}&\hspace{5ex} = \pmatrix{\alpha & \beta \cr \gamma & \delta}
\pmatrix{S_{11} & S_{12} \cr S_{21} & S_{22}}
\pmatrix{\alpha^{*} & \gamma^{*} \cr \beta^{*} & \delta^{*}} ,
\end{eqnarray}
where $C$ and $G$ are the density matrix and the transformation matrix
given in Eq.(\ref{cocy}) and Eq.(\ref{lt22}) respectively.  According
to the basic property of the Lorentz group, these transformations do
not change the determinant of the density matrix $C$.  Transformations
which do not change the determinant are called unimodular
transformations.

As we shall see in this section, the determinant for pure states is
zero, while that of mixed states does not vanish.  Is there then
a transformation matrix which will change this determinant within the
Lorentz group.  The answer is No.  This is the basic issue we would
like to address in this section.

If the phase difference between the two waves remains intact, the
system is said to in a pure state, and the density matrix can be
brought to the form
\begin{equation}\label{pure22}
\pmatrix{1 & 0 \cr 0 & 0} ,
\end{equation}
through the transformation of Eq.(\ref{trans22}) with a suitable
choice of the $G$ matrix.  For the pure state, the Stokes four-vector
takes the form
\begin{equation}\label{pure4}
\pmatrix{1 \cr 1 \cr 0 \cr 0} .
\end{equation}

In order to study the symmetry properties of the density matrix, let
us ask the following question.  Is there a group of transformation
matrices which will leave the above density matrix invariant?
In answering this question, it is more convenient to use the Stokes
four-vector.  The column vector of Eq.(\ref{pure4}) is invariant under
the operation of the phase shifter $P(\phi)$ of Eq.(\ref{shif44}).
In addition, it is invariant under the following two matrices:
\begin{eqnarray}\label{d1d2}
&{}& F_{1}(u) = \pmatrix{ 1 + u^{2}/2  & - u^{2}/2 & u  & 0  \cr
   u^{2}/2  & 1 - u^{2}/2 & u & 0   \cr
   u  & -u & 1 & 0 \cr
   0 & 0 & 0 & 1 } ,  \nonumber \\[2ex]
&{}& F_{2}(v) = \pmatrix{ 1 + v^{2}/2  & - v^{2}/2 & 0  & v  \cr
   v^{2}/2  & 1 - v^{2}/2 & 0 & v   \cr
   0  & 0 & 1 & 0 \cr
   u & -v & 0 & 1 } .
\end{eqnarray}
These mathematical expressions were first discovered by Wigner in
1939~\cite{wig39} in connection with the internal space-time
symmetries of relativistic particles.  They went through a stormy
history, but it is gratifying to note that they serve a useful purpose
for studying interferometers where each matrix corresponds to
an operation which can be performed in laboratories.

The $F_{1}$ and $F_{2}$ matrices commute with each other, and the
multiplication of these leads to the form
\begin{equation}\label{d44}
F_{2}(u)F_{2}(v)
= \pmatrix{1 + (u^{2} + v^{2})/2  & - (u^{2} + v^{2})/2 & u & u \cr
   (u^{2} + v^{2})/2  & 1 - (u^{2} + v^{2})/2 & u & v   \cr
   u  & -u & 1 & 0 \cr
   v & -v & 0 & 1 } .
\end{equation}
\noindent This matrix contains two parameters.

Let us go back to the phase-shift matrix of Eq.(\ref{shif44}).
This matrix also leaves the Stokes vector of Eq.(\ref{pure4})
invariant.  If we define the ``little group'' as the maximal subgroup
of the Lorentz group which leaves a Stokes vector invariant, the
little group for the Stokes vector of Eq.(\ref{pure4}) consists of the
transformation matrices given in Eq.(\ref{shif44}) and Eq.(\ref{d44}).

Next, if the phase relation is completely random, and the first and
second components have the same amplitude, the density matrix becomes
\begin{equation}\label{imp22}
\pmatrix{1/2 & 0 \cr 0 & 1/2} .
\end{equation}
Here is the question: Is there a two-by-two matrix which will
transform the pure-state density matrix of Eq.(\ref{pure22}) into the
impure-state matrix of Eq.(\ref{imp22})?  The answer within the system
of matrices of the form given in Eq.(\ref{lt22}) is No, because the
determinant of the pure-state density matrix is zero while that of
the impure-state matrix is $1/4$.  Is there a way to deal with this
problem?  This problem was addressed in Ref.~\cite{hkn00}.
In this section, we restrict ourselves to the unimodular transformation
of Eq.(\ref{trans22}) which preserves the value of the determinant of
the density matrix.  The Stokes four-vector corresponding to the above
density matrix is
\begin{equation}\label{imp4}
\pmatrix{1 \cr 0 \cr 0 \cr 0} .
\end{equation}
This vector is invariant under both the rotation matrix of
Eq.(\ref{rot44}) and the phase shift matrix of Eq.(\ref{shif44}).
Repeated applications of these matrices lead to a three-parameter
group of rotations applicable to the three-dimensional space of
$(S_{1}, S_{2}, S_{3})$.

Not all the impure-state density matrices take the form of
Eq.(\ref{imp22}).  In general, if they are brought to a diagonal
form, the matrix takes the form
\begin{equation}\label{impp22}
{1 \over 2}\pmatrix{1 + \cos\chi & 0 \cr 0 & 1 - \cos\chi} ,
\end{equation}
and the corresponding Stokes four-vector is
\begin{equation}\label{impp4}
e^{-\eta} \pmatrix{\cosh\eta \cr \sinh\eta \cr 0 \cr 0} ,
\end{equation}
with
\begin{equation}
\eta = {1 \over 2}\ln{1 + \cos\chi \over 1 - \cos\chi} .
\end{equation}
The matrix which transforms Eq.(\ref{imp4}) to Eq.(\ref{impp4})
is the squeeze matrix of Eq.(\ref{sq44}).
The question then is whether it is possible to transform the pure state
of Eq.(\ref{pure4}) to the impure state of Eq.(\ref{impp4}) or to that
of Eq.(\ref{imp4}).

In order to see the problem in terms of the two-by-two density matrix,
let us go back to the pure-state density matrix of Eq.(\ref{pure22}).
Under the rotation of Eq.(\ref{rot22}),
\begin{equation}
\pmatrix{\cos(\chi/2) & -\sin(\chi/2) \cr \sin(\chi/2) &
\cos(\chi/2) }  \pmatrix{1 & 0 \cr 0 & 0}
\pmatrix{\cos(\chi/2) &  \sin(\chi/2) \cr
-\sin(\chi/2) & \cos(\chi/2) } ,
\end{equation}
the pure-state density matrix becomes
\begin{equation}
{1 \over 2} \pmatrix{1 + \cos\chi & \sin\chi \cr
                      \sin\chi & 1 - \cos\chi} .
\end{equation}

For the present case of two-by-two density matrices, the trace of the
matrix is one for both pure and impure cases.  The trace of the
$(matrix)^{2}$ is one for the pure state, while it is less than one for
impure states.

The next question is whether there is a two-by-two matrix which will
eliminate the off-diagonal elements of the above expression that
will also lead to the expression of Eq.(\ref{impp22}).  In order to
answer this question, let us note that the determinant of the density
matrix vanishes for the pure state, while it is non-zero for impure
states.  The Lorentz-like transformations of Eq.(\ref{trans22}) leave
the determinant invariant.  Thus, it is not possible to transform a
pure state into an impure state by means of the transformations from
the six-parameter Lorentz group.  Then is it possible to achieve this
purpose using two-by-two matrices not belonging to this group.  We do
not know the answer to this question.  We are thus forced to resort
to four-by-four matrices applicable to the Stokes four-vector.

\section{Decoherence Effects on the Little Groups}\label{little}
We are interested in a transformation which will change the density
matrix of Eq.(\ref{pure22}) to that of Eq.(\ref{imp22}).  For this
purpose, we can use the Stokes four-vector consisting of the four
elements of the density matrix.  The question then is
whether it is possible to find a transformation matrix which will
transform the pure-state four-vector of Eq.(\ref{pure4}) to the
impure-state four-vector of Eq.(\ref{imp4}).

Mathematically, it is more convenient to ask whether the inverse of
this process is possible: whether it is possible to transform the
four-vector of Eq.(\ref{imp4}) to that of Eq.(\ref{pure4}).  This is
known in mathematics as the contraction of the three-dimensional
rotation group into the two-dimensional Euclidean group~\cite{knp86}.
Let us apply the squeeze matrix of Eq.(\ref{sq44}) to the four-vector
of Eq.(\ref{imp4}).  This can be written as
\begin{equation}\label{sqimp}
\pmatrix{\cosh\eta & \sinh\eta & 0 & 0 \cr
\sinh\eta & \cosh\eta & 0 & 0 \cr
0 & 0 & 1 & 0 \cr 0 & 0 & 0 & 1}
\pmatrix{1 \cr 0 \cr 0 \cr 0} =
\pmatrix{\cosh\eta \cr \sinh\eta \cr 0 \cr 0} .
\end{equation}
After an appropriate normalization, the right-hand side of the above
equation becomes like the pure-state vector of Eq.(\ref{pure4}) in the
limit of large $\eta$, as $\cosh\eta$ becomes equal to $\sinh\eta$
in the infinite-$\eta$ limit.  This transformation is from a mixed
state to a pure or almost-pure state.  Since we are interested in the
transformation from the pure state of Eq.(\ref{pure4})  to the impure
state of Eq.(\ref{imp4}), we have to consider an inverse of the above
equation:
\begin{equation}\label{inver1}
\pmatrix{\cosh\eta & -\sinh\eta & 0 & 0 \cr
-\sinh\eta & \cosh\eta & 0 & 0 \cr 0 & 0 & 1 & 0 \cr 0 & 0 & 0 & 1}
\pmatrix{\cosh\eta \cr \sinh\eta \cr 0 \cr 0}  =
\pmatrix{1 \cr 0 \cr 0 \cr 0} .
\end{equation}
However, the above equation does not start with the pure-state
four-vector.  If we apply the same matrix to the pure state matrix,
the result is
\begin{equation}
\pmatrix{\cosh\eta & -\sinh\eta & 0 & 0 \cr
-\sinh\eta & \cosh\eta & 0 & 0 \cr 0 & 0 & 1 & 0 \cr 0 & 0 & 0 & 1}
\pmatrix{1 \cr 1 \cr 0 \cr 0}
= e^{-\eta} \pmatrix{1 \cr 1 \cr 0 \cr 0} .
\end{equation}
The resulting four-vector is proportional to the pure-state four-vector
and is definitely not an impure-state four-vector.

The inverse of the transformation of Eq.(\ref{sqimp}) is not capable
of bringing the pure-state vector into an impure-state vector.  Let us
go back to Eq.(\ref{sqimp}), it is possible to bring an impure-state
into a pure state only in the limit of infinite $\eta$.  Otherwise,
it is not possible.  It is definitely not possible if we take into
account experimental considerations.

The story is different for the little groups.  Let us start with
the rotation matrix of Eq.(\ref{rot44}), and apply to this matrix
the transformation matrix of Eq.(\ref{sqimp}).  Then
\begin{eqnarray}\label{3mats}
&{}&\pmatrix{\cosh\eta & \sinh\eta & 0 & 0 \cr
  \sinh\eta & \cosh\eta & 0 & 0 \cr
  0 & 0 & 1 & 0 \cr 0 & 0 & 0 & 1}
\pmatrix{1 & 0 & 0 & 0 \cr 0 & \cos\theta & -\sin\theta & 0 \cr
0 & \sin\theta & \cos\theta & 0 \cr 0 & 0 & 0 & 1} \nonumber\\[2ex]
&{}& \times \pmatrix{\cosh\eta & -\sinh\eta & 0 & 0 \cr
  -\sinh\eta & \cosh\eta & 0 & 0 \cr
   0 & 0 & 1 & 0 \cr 0 & 0 & 0 & 1} .
\end{eqnarray}
\noindent If $\eta$ is zero, the above expression becomes the rotation
matrix of Eq.(\ref{rot44}).  If $\eta$ becomes infinite, it becomes
the little-group matrix $F_{1}(u)$ of Eq.(\ref{d1d2}) applicable to
the pure state of Eq.(\ref{pure4}).  The details of this calculation
for the case of Lorentz transformations are given in the 1986 paper
by Han {\it et al.}~\cite{hks86jm}.  We are then led to the question of
whether one little-group transformation matrix can be transformed from
the other.

If we carry out the matrix algebra of Eq.(\ref{3mats}), the result is
\begin{equation}
\pmatrix{1 + \alpha u^{2} w/2 & -\alpha u^{2} w/2 & \alpha uw & 0 \cr
\alpha u^{2} w/2 & 1 - u^{2} w/2 & uw & 0 \cr
\alpha uw & -uw & 1 - (1 - \alpha^{2}) u^{2} w/2 & 0 \cr
0 & 0 & 0 & 1} ,
\end{equation}
where
\begin{equation}\label{anal}
\alpha = \tanh\eta , \qquad
u = - 2\,\tan\left({\theta \over 2}\right), \qquad
w = { 1 \over 1 + (1 - \alpha^{2})\tan^{2}(\theta/2)}.
\end{equation}
\noindent  If $\alpha = 0$, the above expression becomes the rotation
matrix of Eq.(\ref{rot44}).  If $\alpha = 1$, it becomes the $F_{1}$
matrix of Eq.(\ref{d1d2}).  Here we used the parameter $\alpha$
instead of $\eta$.  In terms of this parameter, it is possible to make
an analytic continuation from the pure state with $\alpha = 1$ to an
impure state with $\alpha < 1$ including $\alpha = 0$.

On the other hand, we should keep in mind that the determinant of the
density matrix is zero for the pure state, while it is non-zero for
all impure states.  For $\alpha = 1$, the determinant vanishes, but
it is nonzero and stays the same for all non-zero values of $\alpha$
less than one and greater than or equal to zero.  The analytic
expression of Eq.(\ref{anal}) hides this singular nature of the
little group~\cite{hks86jm}.

\section{Poincar\'e Sphere as the Representation of the
Lorentz group}\label{poinc}

In Secs.~\ref{polari}, \ref{interfer}, and \ref{little}, it was noted
that the Stokes parameters form a four-vector in the Minkowskian space.
Thus, it was possible to discuss the density matrix in terms of the
four-vectors.

The Poincar\'e sphere was originally constructed from polarization
optics.  Therefore, it is also a representation of the Lorentz group.
The Poincar\'e sphere for various polarization states have been
thoroughly discussed in a recent book by Brosseau~\cite{bross98}.

What is interesting is that the Poincar\'e sphere has two radii.
One of them is the maximum radius specified by $S_{0}$, and the
other radius is the length of the space-like components, namely
\begin{equation}
S = \sqrt{S_{1}^{2} + S_{2}^{2} + S_{3}^{2}} .
\end{equation}
According to the four-vector property of the Stokes parameter,
\begin{equation}
M^{2} = S_{0}^{2} - S^{2}
\end{equation}
is invariant under Lorentz transformations.  On the other hand,
$M = 0$ for a fully coherent state, while it is non-zero for a
partially coherent state.  The system is totally incoherent if $S = 0$.
However, the Lorentz group cannot handle this decoherence mechanism.

We observe here that the pure state with $M = 0$ corresponds to the
$E(2)$-like little group for massless particles while it corresponds
to the $O(3)$-like little group for non-zero values of $M$.  The
transition of $O(3)$ to $E(2)$ is well known as the group contraction.
However, the inverse transformation requires further study.

\section{One-lens System}\label{olens}
In analyzing optical rays in para-axial lens optics, we start with
the lens matrix:
\begin{equation}\label{lens}
 L = \pmatrix{1 & 0 \cr -1/f & 1} ,
\end{equation}
and the translation matrix
\begin{equation}\label{trans}
T = \pmatrix{1 & z \cr 0 & 1} ,
\end{equation}
assuming that the beam is propagating along the $z$ direction.

Then the one-lens system consists of
\begin{equation}
\pmatrix{1 & z_{2} \cr 0 & 1}
\pmatrix{1 & 0 \cr -1/f & 1}
\pmatrix{1 & z_{1} \cr 0 & 1} .
\end{equation}
If we perform the matrix multiplication,
\begin{equation}
\pmatrix{1 - z_{2}/f   & z_{1} + z_{2} - z_{1}z_{2}/f   \cr
-1/f & 1 - z_{1}/f } .
\end{equation}
If we assert that the upper-right element be zero,  then
\begin{equation}
{1 \over z_{1}} + {1 \over z_{2}} = {1 \over f} ,
\end{equation}
and the image is focussed, where $z_{1}$ and $z_{2}$ are the distance
between the lens and object and between the lens and image
respectively.  They are in general different, but we shall assume for
simplicity that they are the same: $z_{1} = z_{2} = z$.  We are doing
this because this simplicity does not destroy the main point of our
discussion, and because the case with two different values has been
dealt with in the literature~\cite{gk03}.  Under this assumption, we
are left with
\begin{equation}
\pmatrix{1 - z/f  &  2z - z^{2}/f  \cr  -1/f  & 1 - z/f } .
\end{equation}
The diagonal elements of this matrix are dimensionless.  In order to
make the off-diagonal elements dimensionless, we write this matrix as
\begin{equation}
 - \pmatrix{\sqrt{z}  & 0 \cr 0 & 1/\sqrt{z}}
\pmatrix{1 - z/f &  z/f - 2  \cr  z/f  & 1 - z/f}
\pmatrix{\sqrt{z}  & 0 \cr 0 & 1/\sqrt{z}} .
\end{equation}
Indeed, the matrix in the middle contains dimensionless elements.  The
negative sign in front is purely for convenience.  We are then led to
study the core matrix
\begin{equation}\label{core}
C = \pmatrix{ x - 1  & x -2  \cr x  & x - 1} .
\end{equation}

Here, the important point is that the above matrices can be written
in terms of transformations of the Lorentz group.  In the two-by-two
matrix representation, the Lorentz boost along the $z$ direction takes
the form
\begin{equation}
B(\eta) = \pmatrix{\exp{(\eta/2)} & 0 \cr 0 & \exp{(-\eta/2)} } ,
\end{equation}
and the rotation around the $y$ axis can be written as
\begin{equation}
R(\theta) = \pmatrix{\cos(\phi/2) & -\sin(\phi/2) \cr
\sin(\phi/2) & \cos(\phi/2) } ,
\end{equation}
and the boost along the $x$ axis takes the form
\begin{equation}
X(\chi) = \pmatrix{\cosh(\chi/2) & \sinh(\chi/2) \cr
\sinh(\chi/2) & \cosh(\chi/2)} .
\end{equation}
Then the core matrix of Eq.(\ref{core}) can be written as
\begin{equation}\label{phi11}
B(\eta) R(\phi) B(-\eta),
\end{equation}
or
\begin{equation}\label{phi22}
\pmatrix{\cos(\phi/2)  &
- e^{-\eta}\sin(\phi/2) \cr  e^{\eta}\sin(\phi/2) & \cos(\phi/2) } ,
\end{equation}
if $ 1 < x < 2 $.  If $x$ is greater than 2, the upper-right element
of the core is positive and it can take the form
\begin{equation}\label{chi11}
B(\eta) X(\chi) B(-\eta),
\end{equation}
or
\begin{equation}\label{chi22}
 \pmatrix{\cosh(\chi/2)  &
 e^{-\eta}\sinh(\chi/2) \cr  e^{+\eta}\sinh(\chi/2) & \cosh(\chi/2) } .
\end{equation}

The expressions of Eq.(\ref{phi11}) and Eq.(\ref{chi11}) are a Lorentz
boosted rotation and a Lorentz-boosted boost matrix along the $x$
direction respectively.  These expressions play the key role in
understanding Wigner's little groups for relativistic particles.

Let us look at their explicit matrix representations given in
Eq.(\ref{phi22}) and Eq.(\ref{chi22}).  The transition from
Eq.(\ref{phi22}) to Eq.(\ref{chi22}) requires the upper right element
going through zero.  This can only be achieved through $\eta$ going to
infinity.  If we like to keep the lower-left element finite during
this process, the angle $\phi$ and the boost parameter $\chi$ have to
approach zero.  The process of approaching the vanishing upper-right
element is necessarily a singular transformation.  This aspect plays
the key role in unifying the internal space-time symmetries of massive
and massless particles.  This is like Einstein's
$E = \sqrt{(pc)^{2} + m_{0}^{2}c^{4}}$
becoming $E = pc$ in the limit of large momentum.

On the other hand, the core matrix of Eq.(\ref{core}) is an analytic
function of the variable $x$.  Thus, the lens matrix allows a
parametrization which allows the transition from massive particle to
massless particle analytically.  The lens optics indeed serves as the
analogue computer for this important transition in particle physics.

>From the mathematical point of view,  Eq.(\ref{phi22}) and
Eq.(\ref{chi22}) represent circular and hyperbolic geometries,
respectively.  The transition from one to the other is not a trivial
mathematical procedure.  It requires a further investigation.

Let us go back to the core matrix of Eq.(\ref{core}).  The $x$
parameter does not appear to be a parameter of Lorentz transformations.
However, the matrix can be written in terms of another set of Lorentz
transformations. This aspect has been discussed in the
literature~\cite{bk03}.

\section{Multi-lens Problem}\label{mlens}
Let us consider a co-axial system of an arbitrary number of lens.
Their focal lengths are not necessarily the same, nor are their
separations. We are then led to consider an arbitrary number of the
lens matrix given in Eq.(\ref{lens}) and an arbitrary number of
translation matrix of Eq.(\ref{trans}).  They are multiplied like
\begin{equation}\label{lsystem}
T_{1}~ L_{1}~T_{2}~ L_{2}~T_{3}~ L_{3}............T_{N}~ L_{N} ,
\end{equation}
where $N$ is the number of lenses.

The easiest way to tackle this problem in to use the Lie-algebra
approach.  Let us start with the generators of the Sp(2) group:
\begin{eqnarray}\label{sq11}
B_{1} = { 1 \over 2}\pmatrix{i  & 0 \cr 0 & -i }, \qquad
B_{2} = { 1 \over 2}\pmatrix{0  & i \cr i & 0 }, \qquad
J = { 1 \over 2}\pmatrix{0  & -i \cr i & 0 } .
\end{eqnarray}
Since the generators are pure imaginary, the transformation matrices
are real.

On the other hand, the $L$ and $T$ matrices of Eq.(\ref{lens}) and
Eq.(\ref{trans}) are generated by
\begin{equation}
X_{1} = \pmatrix{ 0 & i \cr 0 & 0},  \qquad
X_{2} = \pmatrix{ 0 & 0 \cr i & 0} .
\end{equation}
If we introduce the third matrix
\begin{equation}
X_{3} = \pmatrix{i & 0 \cr 0 & -i} ,
\end{equation}
all three matrices form a closed set of commutation relations:
\begin{equation}\label{shear}
\left[X_{1}, X_{2}\right] = iX_{3}, \qquad
\left[X_{1}, X_{3}\right] = -iX_{1},  \qquad
\left[X_{2}, X_{3}\right] = iX_{2} .
\end{equation}
Thus, these generators also form a closed set of Lie algebra generating
real two-by-two matrices.  What group would this generate?  The answer
has to be $Sp(2)$.  The truth is that the three generators given
in Eq.(\ref{shear}) can be written as linear combinations of the
generators of the $Sp(2)$ group given in Eq.(\ref{sq11})~\cite{bk01}.
Thus, the $X_{i}$ matrices given above can also act as the generators
of the $Sp(2)$ group, and the lens-system matrix given in
Eq.(\ref{lsystem}) is a three-parameter matrix of the form of
Eq.(\ref{alpha2}) with real elements.

The resulting real matrix is written as
\begin{equation}
\pmatrix{A & B \cr C & D} ,
\end{equation}
and is called the $ABCD$ matrix.  The question then is how many
lenses are need to give the most general form of the $ABCD$
matrix~\cite{sudar85}.  This matrix has three independent parameter.

According to Bargmann~\cite{barg47}, this three-parameter matrix can
be decomposed into
\begin{equation}
\pmatrix{\cos(\alpha/2)  & -\sin(\alpha/2)
\cr \sin(\alpha/2)  & \cos(\alpha/2)}
\pmatrix{e^{\gamma/2} & 0 \cr  0 & e^{-\gamma/2} }
\pmatrix{\cos(\beta/2)  & -\sin(\beta/2)  \cr
\sin(\beta/2)  & \cos(\beta/2)},
\end{equation}
which can be written as the product of one symmetric matrix resulting
from
\begin{equation} \label{symm}
\pmatrix{\cos(\alpha/2)  & -\sin(\alpha/2)
\cr \sin(\alpha/2)  & \cos(\alpha/2)}
\pmatrix{e^{\gamma/2} & 0 \cr  0 & e^{-\gamma/2} }
\pmatrix{\cos(\alpha/2)  & \sin(\alpha/2)  \cr
-\sin(\alpha/2)  & \cos(\alpha/2)}
\end{equation}
and one rotation matrix:
\begin{equation}\label{ortho}
\pmatrix{\cos[(\beta + \alpha)/2]  & -\sin[(\beta + \alpha)/2]   \cr
\sin[(\beta + \alpha)/2]   & \cos[(\beta + \alpha)/2] } .
\end{equation}
The expression in Eq.(\ref{symm}) is a symmetric matrix, while that of
Eq.(\ref{ortho}) is an orthogonal one.  We can then decompose each of
these two matrices into the lens and translation matrices.  The net
result is that we do not need more than three lenses to describe the
lens system consisting of an arbitrary number of lenses.  The detailed
calculations are given in Ref.~\cite{bk01}.

\section{Laser Cavities}\label{lcav}
In a laser cavity, the optical ray makes round trips between two
mirrors.  One cycle is therefore equivalent to a two-lens system with
two identical lenses and the same  distance between the lenses.  Let
us rewrite the matrix corresponding to the one-lens system given in
Eq.(\ref{core}).
\begin{equation}\label{core22}
C = \pmatrix{ x - 1  & x -2  \cr x  & x - 1} .
\end{equation}
Then one complete cycle consists of $C^{2}$.

Here, the cycle starts from the midpoint between the two
mirrors~\cite{bk01}, unlike the traditional approach to this problem
where the cycle starts from the surface of one of the two
mirrors~\cite{haus84}.  By choosing the midpoint, we can eliminate
the auxiliary flat mirror between them needed in the traditional
approach. This problem was discussed in detail in Ref.~\cite{bk01}.

For $N$ cycles, the expression should be $C^{2N}$.  This calculation,
using the above expression, will not lead to a manageable form.
However, we can resort to the expressions of Eq.(\ref{phi11}) and
Eq.(\ref{chi11}). Then one cycle consists of
\begin{eqnarray}\label{cycle}
&{}& C^{2} = [B(\eta) R(\phi/2) B(-\eta)]
[B(\eta) R(\phi/2) B(-\eta)]    \nonumber \\[2ex]
&{}& \hspace{5mm} = B(\eta) R(\phi) B(-\eta) ,
\end{eqnarray}
if the upper-right element is negative.  If it is positive, the
expression should be
\begin{eqnarray}
&{}& C^{2} = [B(\eta) X(\chi/2) B(-\eta)]
[B(\eta) X(\chi/2) B(-\eta)]    \nonumber \\[2ex]
&{}& \hspace{5mm} = B(\eta) X(\chi) B(-\eta) .
\end{eqnarray}

If these expressions are repeated $N$ times,
\begin{equation}\label{stable}
C^{2N} = B(\eta) R(N\phi) B(-\eta) ,
\end{equation}
if the upper-right element is negative.  If it is positive, the
expression should be
\begin{equation}
C^{2N} = B(\eta) X(N\chi) B(-\eta) .
\end{equation}
As $N$ becomes large,  $\cosh(N\chi)$ and $\sinh(N\chi)$ become very
large, the beam deviates from the laser cavity.   Thus, we have to
restrict ourselves to the case given in Eq.(\ref{stable}).

The core of the expression of Eq.(\ref{stable}) is the rotation
matrix
\begin{equation}
       R(N\phi) = [R(\phi)]^{N}     .
\end{equation}
This means that one complete cycle in the cavity corresponds to the
rotation matrix $R(\phi)$.  The rotation continues as the beam
continues to repeat the cycle.

Let us go back to Eq.(\ref{cycle}).  The expression corresponds to
a Lorentz boosted rotation, or bringing a moving particle to its rest
frame, rotate, and boost back to the original momentum.  The rotation
associated with the momentum-preserving transformation is called
the Wigner's little-group rotation, which is related to the Wigner
rotation commonly mentioned in the literature~\cite{bk01}.

\section{Multi-layer Optics}\label{mlayer}
The most efficient way to study multi-layer optics is to use the
S-matrix formalism~\cite{azzam77}.  This formalism is also based on
two-by-two matrices, and we can write
\begin{equation}
\pmatrix{\psi_{1} \cr \psi_{2}} = \pmatrix{A & B \cr C  & D}
\pmatrix{\psi_{3} \cr 0} ,
\end{equation}
where $\psi_{1}, \psi_{2}$ and $\psi_{3}$ are the incoming, reflected
and transmitted beams.  Here we use the matrix notion $ABCD$ for the
two-by-two $S$ matrix.

We consider a system of two different optical layers.  For convenience,
we start from the boundary from $medium~2$ to $medium~1$.
We can write the boundary matrix as~\cite{monzon96}
\begin{equation}\label{bd11}
B(\eta) = \pmatrix{\cosh(\eta/2) & \sinh(\eta/2) \cr \sinh(\eta/2) &
\cosh(\eta/2) } ,
\end{equation}
taking into account both the transmission and reflection of the beam.
The parameter $\eta$ is of course determined by the transmission and
reflection coefficient at the surface.  This problem was studied in
depth by Monzon {\it et al.}~\cite{monzon96}, and their research on
this subject is still continuing.

As the beam goes through the $medium~1$, the beam undergoes the phase
shift represented by the matrix
\begin{equation}\label{ps11}
P(\phi_1) = \pmatrix{e^{-i\phi_1/2} & 0 \cr 0 & e^{i\phi_1/2}} .
\end{equation}
When the wave hits the surface of the second medium, the boundary
matrix is
\begin{equation}\label{bd22}
B(-\eta) = \pmatrix{\cosh(\eta/2) & -\sinh(\eta/2) \cr -\sinh(\eta/2)
& \cosh(\eta/2) } ,
\end{equation}
which is the inverse of the matrix given in Eq.(\ref{bd11}).
Within the second medium, we write the phase-shift matrix as
\begin{equation}\label{ps22}
P(\phi_2) = \pmatrix{e^{-i\phi_2/2} & 0 \cr 0 & e^{i\phi_2/2}} .
\end{equation}
Then, when the wave hits the first medium from the second, we have
to go back to Eq.(\ref{bd11}).
Thus, the one cycle consists of
\begin{eqnarray}\label{m1}
&{}& \pmatrix{\cosh(\eta/2) &
\sinh(\eta/2) \cr \sinh(\eta/2) & \cosh(\eta/2) }
\pmatrix{e^{-i\phi_1/2} & 0 \cr 0 & e^{i\phi_1/2}}\nonumber \\[2ex]
&{}& \hspace{10mm}
\times \pmatrix{\cosh(\eta/2) & -\sinh(\eta/2)
\cr -\sinh(\eta/2) & \cosh(\eta/2) }
\pmatrix{e^{-i\phi_2/2} & 0 \cr 0 & e^{i\phi_2/2}} .
\end{eqnarray}

The above matrices contain complex numbers.  However, it is possible
to transform simultaneously
\begin{equation}
\pmatrix{\cosh(\eta/2) & \sinh(\eta/2) \cr \sinh(\eta/2) &  \cosh(\eta/2)}
\end{equation}
to
\begin{equation}
\pmatrix{\exp{(\eta/2)} & 0 \cr 0 & \exp{(-\eta/2)} },
\end{equation}
and transform
\begin{equation}
\pmatrix{e^{-i\phi_i/2} & 0 \cr 0 & e^{i\phi_i/2}}
\end{equation}
to
\begin{equation}
\pmatrix{\cos(\phi_i/2) & -\sin(\phi_i/2) \cr
\sin(\phi_i/2)  & \cos(\phi_i/2) } ,
\end{equation}
using a conjugate transformation, as is shown in Appendix~\ref{conju}.
It is also possible to transform these expressions back to their
original forms.  This transformation property has been discussed in
detail in Ref.~\cite{gk01}, and also in Appendix~\ref{conju} of the
present paper.

As a consequence, the matrix of Eq.(\ref{m1}) becomes
\begin{eqnarray}\label{core55}
&{}&\pmatrix{e^{\eta/2} & 0 \cr 0 & e^{-\eta/2} }
\pmatrix{\cos(\phi_1/2) & -\sin(\phi_1/2) \cr
\sin(\phi_1/2) &  \cos(\phi_1/2)}   \nonumber \\[2ex]
&{}&  \hspace{10mm}\times \pmatrix{ e^{-\eta/2} & 0 \cr
0 & e^{\eta/2}}
\pmatrix{\cos(\phi_2/2) & -\sin(\phi_2/2) \cr
\sin(\phi_2/2) &  \cos(\phi_2/2) }  .
\end{eqnarray}
In the above expression, the first three matrices are of the same
mathematical form as that of the core matrix for the one-lens system
given in Eq.(\ref{phi11}).  The fourth matrix is an additional rotation
matrix.  This makes the mathematics of repetition more complicated, but
this has been done~\cite{gk03}.

As a consequence the net result becomes
\begin{equation}
      B(\mu) R(N\alpha) B(-\mu) ,
\end{equation}
or
\begin{equation}
      B(\mu) X(N\xi) B(-\mu),
\end{equation}
where the parameters $\mu, \alpha$ and $\xi$ are to be determined
from the input parameters $\eta, \phi_1$ and $\phi_2$.  Detailed
calculations are given in Ref.~\cite{gk03}.

It is interesting to note that the Lorentz group can serve as
a computational device also in multi-layer optics.

\section{Concluding Remarks}
We have seen in this report that the Lorentz group provides convenient
calculational tools in many branches of ray optics.  The reason is that
ray optics is largely based on two-by-two matrices.   These matrices
also constitute the group $SL(2,c)$ which serves as the universal
covering group of the Lorentz group.

The optical instruments discussed in this report are the fundamental
components in optical circuits.  In the world of electronics, electric
circuits form the fabric of the system.  In the future high-technology
world, optical components will hold the key to technological
advances.  Indeed, the Lorentz group is the fundamental language for
the new world.

It is by now well known that the Lorentz group is the basic language
for quantum optics.  Coherent and squeezed states are representations
of the Lorentz group. It is challenging to see how the Lorentz nature
of the above-mentioned optical components will manifest itself in
quantum world.

The Lorentz group was introduced to physics by Einstein and Wigner to
understand the space-time symmetries of relativistic particles and
the covariant world of electromagnetic fields. It is gratifying to
note that the Lorentz group can serve as the language common both to
particle physics and optical sciences.

\section*{Appendix}

\appendix

\section{Lorentz-squeezed  Harmonic Oscillators}\label{sqosc}
The purpose of this Appendix is to expand the third row of
Table~\ref{einwig} using the covariant formalism of harmonic
oscillators.  We explain first what this formalism does for covariant
picture of relativistic extended particles.  We then point out that
this oscillator formalism can serve as the basic language for two-mode
coherent states~\cite{yuen76}, namely squeezed states of light.

The concept of localized probability distribution is the backbone of
the present formulation of quantum mechanics.  Of course, we would
like to have more deterministic form of dynamics, and efforts have
been and are still being made along this direction.  One of the most
serious problems with this probabilistic interpretation is whether
this concept of probability is consistent with special relativity.

In a given Lorentz frame, we know how to do quantum mechanics with
a localized probability distribution.  How would this distribution
look to an observer in a different Lorentz frame?

\begin{itemize}
\item  Would the probability distribution appear the same to this
      observer?

\item  If different, how is the probability distribution distorted?

\item  Is the total probability conserved?

\item  The Lorentz boost mixes up the spatial coordinate with the
       time variable.  What role does the time-separation variable
       play in defining the boundary condition for localization and
       the probability distribution?
\end{itemize}

We can answer some or all of the above questions only if we construct
covariant wave functions, namely wave functions which can be
Lorentz-boosted.  It is easy to construct these wave functions if we
know the answer.  In the initial development of quantum mechanics,
the harmonic oscillator played the pivotal role.  Thus, it is clear
to us that if there is a wave function which can be Lorentz-boosted,
this has to be the harmonic oscillator wave function.  Until we
construct wave functions which can be Lorentz-transformed, we cannot
say that quantum mechanics is consistent with relativity.  Indeed,
we should examine this problem before attempting to invent more
definitive quantum mechanics.

Since the hadron, in the quark model, is a bound-state of
quarks~\cite{gell64}, Feynman, Kislinger, and Ravndal, in
1971~\cite{fkr71}, raised the following question in connection with
the quark model for hadrons: the hadronic spectrum can indeed be
explained in terms of the degeneracy of three-dimensional harmonic
oscillator wave functions in the hadronic rest frame; however, what
happens when the hadron moves? Indeed, Feynman
{\it et al.} wrote a Lorentz-invariant differential equation whose
solutions can become non-relativistic wave functions if the
time-separation variable can be ignored.

This Lorentz-invariant differential equation is a four-dimensional
partial differential equation, with many different solutions depending
on separation of variables and boundary conditions.  There is a set
of normalizable solutions which can serve as a representation space
for Wigner's little group for massive particles~\cite{wig39,knp86}.
We can start with this set of solutions and give physical
interpretations, especially to the time-separation variable.

Finally, is the covariance of the oscillator wave function consistent
with what we observe in the real world.  Here again, Feynman plays
the key role.  While the quark model can be fit into the oscillator
scheme in the hadronic rest frame, Feynman in 1969 came up with the
idea of partons~\cite{fey69}.  According to Feynman's parton model,
the hadron consists of an infinite number of partons when it moves
with a velocity close to that of light.  Quarks and partons are
believed to be the same particles, but their properties are totally
different.  While the quarks inside the hadron interact coherently
with external signals, partons interact incoherently.  If the partons
are Lorentz-boosted quarks, does the Lorentz boost destroy the
coherence?

Before 1964~\cite{gell64}, the hydrogen atom was used for illustrating
bound states.  These days, we use hadrons which are bound states of
quarks.  Let us use the simplest hadron consisting of two quarks bound
together with an attractive force, and consider their space-time
positions $x_{a}$ and $x_{b}$, and use the variables
\begin{equation}
X = (x_{a} + x_{b})/2 , \qquad x = (x_{a} - x_{b})/2\sqrt{2} .
\end{equation}
The four-vector $X$ specifies where the hadron is located in space and
time, while the variable $x$ measures the space-time separation between
the quarks.  According to Einstein, this space-time separation contains
a time-like component which actively participates as can be seen from
\begin{equation}\label{boostm}
\pmatrix{z' \cr t'} = \pmatrix{\cosh \eta & \sinh \eta \cr
\sinh \eta & \cosh \eta } \pmatrix{z \cr t} ,
\end{equation}
when the hadron is boosted along the $z$ direction.
In terms of the light-cone variables defined as~\cite{dir49}
\begin{equation}
u = (z + t)/\sqrt{2} , \qquad v = (z - t)/\sqrt{2} ,
\end{equation}
the boost transformation of Eq.(\ref{boostm}) takes the form
\begin{equation}\label{lorensq}
u' = e^{\eta } u , \qquad v' = e^{-\eta } v .
\end{equation}
The $u$ variable becomes expanded while the $v$ variable becomes
contracted.

Does this time-separation variable exist when the hadron is at rest?
Yes, according to Einstein.  In the present form of quantum mechanics,
we pretend not to know anything about this variable.  Indeed, this
variable belongs to Feynman's rest of the universe.  In this report,
we shall see the role of this time-separation variable in the
decoherence mechanism.

Also in the present form of quantum mechanics, there is an uncertainty
relation between the time and energy variables.  However, there are
no known time-like excitations.  Unlike Heisenberg's uncertainty
relation applicable to position and momentum, the time and energy
separation variables are c-numbers, and we are not allowed to write
down the commutation relation between them.

How does this space-time asymmetry fit into the world of
covariance~\cite{kn73}.  The answer is that Wigner's $O(3)$-like
little group is not a Lorentz-invariant symmetry, but is a covariant
symmetry~\cite{wig39}.  It has been shown that the time-energy
uncertainty relation applicable to the time-separation variable fits
perfectly into the $O(3)$-like symmetry of massive relativistic
particles~\cite{knp86}.

The c-number time-energy uncertainty relation allows us to write down
a time distribution function without excitations~\cite{knp86}.  If we
use Gaussian forms for both space and time distributions, we can start
with the expression
\begin{equation}\label{ground}
\left({1 \over \pi} \right)^{1/2}
\exp{\left\{-{1 \over 2}\left(z^{2} + t^{2}\right)\right\}}
\end{equation}
for the ground-state wave function.  What do Feynman {\it et al.}
say about this oscillator wave function?

\begin{figure}
\centerline{\includegraphics[scale=0.5]{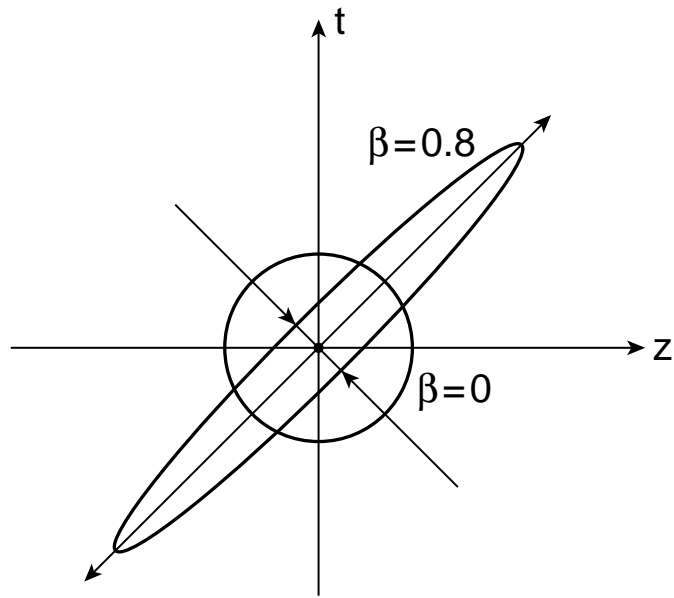}}
\caption{Effect of the Lorentz boost on the space-time
wave function.  The circular space-time distribution at the rest frame
becomes Lorentz-squeezed to become an elliptic distribution.
The first version of this figure consists of two ellipses in the
1973 paper by Kim and Noz~\cite{kn73}.  This figure consisting of a
circle and a ellipse is from the 1978 paper by Kim, Noz, and
Oh~\cite{kno79}.}\label{ellipse}
\end{figure}

In their classic 1971 paper~\cite{fkr71}, Feynman {\it et al.} start
with the following Lorentz-invariant differential equation.
\begin{equation}\label{osceq}
{1\over 2} \left\{x^{2}_{\mu} -
{\partial^{2} \over \partial x_{\mu }^{2}}
\right\} \psi(x) = \lambda \psi(x) .
\end{equation}
This partial differential equation has many different solutions
depending on the choice of separable variables and boundary conditions.
Feynman {\it et al.} insist on Lorentz-invariant solutions which are
not normalizable.  On the other hand, if we insist on normalization,
the ground-state wave function takes the form of Eq.(\ref{ground}).
It is then possible to construct a representation of the Poincar\'e
group from the solutions of the above differential
equation~\cite{knp86}.  If the system is boosted, the wave function
becomes
\begin{equation}\label{eta}
\psi_{\eta }(z,t) = \left({1 \over \pi }\right)^{1/2}
\exp\left\{-{1\over 2}\left(e^{-2\eta }u^{2} +
e^{2\eta}v^{2}\right)\right\} .
\end{equation}
This wave function becomes Eq.(\ref{ground}) if $\eta$ becomes zero.
The transition from Eq.(\ref{ground}) to Eq.(\ref{eta}) is a
squeeze transformation.  The wave function of Eq.(\ref{ground}) is
distributed within a circular region in the $u v$ plane, and thus
in the $z t$ plane.  On the other hand, the wave function of
Eq.(\ref{eta}) is distributed in an elliptic region with the light-cone
axes as the major and minor axes respectively.  If $\eta$ becomes very
large, the wave function becomes concentrated along one of the
light-cone axes.  Indeed, the form given in Eq.(\ref{eta}) is a
Lorentz-squeezed wave  function.  This Lorentz-squeeze mechanism is
illustrated in Fig.~\ref{ellipse}.

There are many different solutions of the Lorentz invariant
differential equation of Eq.(\ref{osceq}).  The solution given in
Eq.(\ref{eta}) is not Lorentz invariant but is covariant.  It is
normalizable in the $t$ variable, as well as in the space-separation
variable $z$.  How can we extract probability interpretation from this
covariant wave function?  This issue has been discussed thoroughly in
the literature~\cite{knp86}.


Another pressing problem in physics is that hadrons, like the proton,
can be regarded as quantum bound states of quarks when they move slowly.
However, they appear like collections of partons when they move with
velocity very close to that of light~\cite{fey69}.  Since the quarks and
partons have quite different properties, it is a challenging problem in
modern physics to show that they are the same but appear different
depending on the observer's Lorentz frame.

It has been shown that this Lorentz-squeezed wave function can explain
the quark model and the parton model as two different manifestations
of one covariant entity~\cite{kn77par}.  Thus, the covariant harmonic
oscillator can occupy the third row of Table~\ref{einwig}.

The Lorentz-squeezed wave function can be written as
\begin{equation}
\psi_{\eta}(z, t) = {1 \over \sqrt{\pi}}
\exp\left\{- {1\over 4}\left[e^{\eta}(z - t)^{2} +
e^{-\eta}(z + t)^{2}\right]\right\} .
\end{equation}

As was discussed in the literature for several different
purposes~\cite{knp91,knp86,kno79aj}, this wave function can be
expanded as
\begin{equation}\label{expan}
\psi_{\eta }(z,t) = {1 \over \cosh\eta}\sum^{}_{k}
\left(\tanh{\eta \over 2}\right)^{k} \phi_{k}(z) \phi_{k}(t) .
\end{equation}
This is an expansion in terms of the oscillations along two different
directions.  If those directions are $x_{1}$ and $x_{2}$, we can write
Eq.(\ref{expan}) as
\begin{equation}
\psi_{\eta }(x_{1},x_{2}) = {1 \over \cosh\eta}\sum^{}_{k}
\left(\tanh{\eta \over 2}\right)^{k} \phi_{k}(x_{1}) \phi_{k}(x_{2}) .
\end{equation}
The mathematics of harmonic oscillators can be translated into
second-quantized Fock space.  In the Fock space, this expression
describes the two-photon coherent state discussed first by Yuen in
1976~\cite{yuen76}.

\section{Euler versus Lie Representations}\label{euler}
The group $SL(2,c)$ consists of two-by-two unimodular matrices whose
elements are complex.  There are therefore six independent parameters,
and thus six generators of the Lie algebra.  This group is locally
isomorphic to the six-parameter Lorentz group or $O(3,1)$ applicable
to the Minkowskian space of three space-like directions and one
time-like direction.

We can construct representations of the group starting from the Lie
algebra given in Secs.~\ref{lorentz} and \ref{spinor}.  We can
construct the representation by  using a method similar to what
Goldstein did for the three-dimensional rotation group in terms of
the Euler angles~\cite{gold80}.  There are three-generators for the
rotation group, but Goldstein starts with rotations around the $z$
and $x$ directions.  Rotations around the $y$ axis and the most
general form for the rotation matrix can be constructed from
repeated applications of those two starting matrices.  Let us call
this type of approach the ``Euler construction.''

In constructing the Lorentz group, we observe first that we need
rotations around two different directions.  As for the boost, we need
boosts along one given direction since boots along other directions
can be achieved by rotations.

There are three basic advantages of this approach.  First, the number
of ``starter'' matrices is less than the number of generators.
For example, we need only two starters for the three-parameter
rotation group.  In our case, we started with two matrices for the
three-parameter group $Sp(2)$ and also for $SU(1,1)$.
Second, each starter matrix takes a simple form and has its own
physical interpretation.

The third advantage can be stated in the following way.  Repeated
applications of the starter matrices will lead to a very complicated
expression.  However, the complicated expression can be decomposed
into the minimum number of starter matrices.  For example, this
number is three for the three-dimensional rotation group.  This
number is also three for $SU(2)$ and $Sp(2)$.  We call this the
Euler decomposition.  The present paper is based on both the Euler
construction and the Euler decomposition.

Among the several useful Euler decompositions, the Iwasawa
decomposition plays an important role in the Lorentz group.   We have
seen in this paper what the decomposition does to the two-by-two
matrices of $Sp(2)$, but it has been an interesting subject since
Iwasawa's first publication on this subject~\cite{iwa49}.  It is beyond
the scope of this paper to present a historical review of the subject.
However, we would like to point out that there are areas of physics
where this important mathematical theorem was totally overlooked.

For instance, in particle theory, Wigner's little groups dictate the
internal space-time symmetries of massive and massless particles which
are locally isomorphic to $O(3)$ and $E(2)$ respectively~\cite{wig39}.
The little group is the maximal subgroup of the Lorentz group whose
transformations do not change the four-momentum of a given
particle~\cite{hkn00}.  The $E(2)$-like subgroup for massless
particles is locally isomorphic to the subgroup of $SL(2,c)$ which
can be started from one of the matrices in Eq.(\ref{shear1}) and the
diagonal matrix of Eq.(\ref{ps11}).  Thus there was an underlying
Iwasawa decomposition while the $E(2)$-like subgroup was decomposed
into rotation and boost matrices~\cite{hks87jm}, but the authors did
not know this.

In optics, there are two-by-two matrices with one vanishing
off-diagonal element.  It was generally known that this has something
to do with the Iwasawa effect, but Simon and Mukunda~\cite{simon98}
and Han {\it et al.}~\cite{hkn99} started treating the Iwasawa
decomposition as the main issue in their papers on polarized light.

In para-axial lens optics, the matrices of the form given in
Eq.(\ref{shear1}) are the starters, and repeated applications of those
two starters will lead to the most general form of $Sp(2)$ matrices.
It had been a challenging problem since 1985~\cite{sudar85} to write
the most general two-by-two matrix in lens optics in terms the minimum
number of those starter matrices.  This problem has been solved
recently~\cite{bk01}, and the central issue in the problem was the
Iwasawa decomposition.

\section{Conjugate Transformations}\label{conju}
The core matrix of Eq.(\ref{core}) contains the chain of the matrices
\begin{equation}\label{su11}
W =\pmatrix{e^{-i\phi} & 0 \cr 0 & e^{i\phi}}
\pmatrix{\cosh\eta & \sinh\eta \cr \sinh\eta & \cosh\eta}
\pmatrix{e^{-i\xi} & 0 \cr 0 & e^{i\xi}}  .
\end{equation}
The Lorentz group allows us to simplify this expression under certain
conditions.

For this purpose, we transform the above expression into a more
convenient form, by taking the conjugate of each of the matrices with
\begin{equation}
C_{1} =  {1 \over \sqrt{2}} \pmatrix{1 & i \cr i & 1} .
\end{equation}
Then $C_{1} W C_{1}^{-1}$ leads to
\begin{equation}\label{sp2}
\pmatrix{\cos\phi & -\sin\phi \cr \sin\phi & \cos\phi}
\pmatrix{\cosh\eta & \sinh\eta \cr \sinh\eta & \cosh\eta}
\pmatrix{\cos\xi & -\sin\xi \cr \sin\xi & \cos\xi} .
\end{equation}
In this way, we have converted $W$ of Eq.(\ref{su11}) into a real
matrix, but it is not simple enough.

Let us take another conjugate with
\begin{equation}
C_{2} =  {1 \over \sqrt{2}} \pmatrix{1 & 1 \cr -1 & 1} .
\end{equation}
Then the conjugate $C_{2} C_{1} W C_{1}^{-1} C_{2}^{-1} $ becomes
\begin{equation}\label{abcd}
\pmatrix{\cos\phi & -\sin\phi \cr \sin\phi & \cos\phi}
\pmatrix{e^{\eta} &  0 \cr 0 & e^{-\eta}}
\pmatrix{\cos\xi & -\sin\xi \cr \sin\xi & \cos\xi} .
\end{equation}
The combined effect of $C_{2}C_{1}$ is
\begin{equation}\label{ccc}
C = C_{2}C_{1} = {1 \over \sqrt{2}} \pmatrix{e^{i\pi/4} &  e^{i\pi/4}
\cr -e^{-i\pi/4} & e^{-i\pi/4}} ,
\end{equation}
with
\begin{equation}
C^{-1} = {1 \over \sqrt{2}} \pmatrix{e^{-i\pi/4} &  -e^{i\pi/4} \cr
e^{-i\pi/4} & e^{i\pi/4}} .
\end{equation}

After multiplication, the matrix of Eq.(\ref{abcd}) will take the
form
\begin{equation}
V = \pmatrix{A &  B \cr C & D} ,
\end{equation}
where $A, B, C,$ and $D$ are real numbers.  If $B$ and $C$ vanish,
this matrix will become diagonal, and the problem will become too
simple.  If, on the other hand, only one of these two elements become
zero, we will achieve a substantial mathematical simplification and
will be encouraged to look for physical circumstances which will lead
to this simplification.

Let us summarize.  We started in this section with the matrix
representation $W$ given in Eq.(\ref{su11}).  This form can be
transformed into the $V$ matrix of Eq.(\ref{abcd}) through the
conjugate transformation
\begin{equation}\label{conju11}
V = C W C^{-1} ,
\end{equation}
where $C$ is given in Eq.(\ref{ccc}).  Conversely, we can recover the
$W$ representation by
\begin{equation}\label{conj22}
W = C^{-1} V C .
\end{equation}
For calculational purposes, the $V$ representation is much easier
because we are dealing with real numbers.  On the other hand, the $W$
representation is of the form for the S-matrix we intend to compute.
It is gratifying to see that they are equivalent.

Let us go back to Eq.(\ref{abcd}) and consider the case where the
angles $\phi$ and $\xi$ satisfy the following constraints.
\begin{equation}\label{2angs}
\phi + \xi = 2\theta, \qquad \phi - \xi = \pi/2 ,
\end{equation}
thus
\begin{equation}\label{phixi}
\phi =  \theta + \pi/4,  \qquad   \xi = \theta - \pi/4 .
\end{equation}
Then in terms of $\theta$, we can reduce the matrix of Eq.(\ref{abcd})
to the form
\begin{equation}\label{abcd2}
\pmatrix{(\cosh\eta)\cos(2\theta)  &
       \sinh\eta - (\cosh\eta)\sin(2\theta)  \cr
       \sinh\eta + (\cosh\eta)\sin(2\theta)  &
  (\cosh\eta)\cos(2\theta) } .
\end{equation}
Thus the matrix takes a surprisingly simple form if the parameters
$\theta$ and $\eta$ satisfy the constraint
\begin{equation}\label{constr}
\sinh\eta = (\cosh\eta)\sin(2\theta) .
\end{equation}\
Then the matrix becomes
\begin{equation}\label{decom2}
\pmatrix{1  &  0  \cr 2\sinh\eta & 1 } .
\end{equation}
This aspect of the Lorentz group is known as the Iwasawa
decomposition~\cite{iwa49}, and has been discussed in the optics
literature~\cite{hkn99,gk01,simon98}.

Matrices of this form are not so strange in optics.  In para-axial
lens optics, the translation and lens matrices are written as
\begin{equation}\label{shear1}
\pmatrix{1 & u \cr 0 & 1} , \quad and  \quad \pmatrix{1 & 0 \cr u & 1} ,
\end{equation}
respectively.  These matrices have the following interesting mathematical
property~\cite{hkn97},
\begin{equation}
\pmatrix{1 & u_{1} \cr 0 & 1} \pmatrix{1 & u_{2} \cr 0 & 1} =
\pmatrix{1 & u_{1} + u_{2} \cr 0 & 1} ,
\end{equation}
and
\begin{equation}
\pmatrix{1 & 0 \cr u_{1} & 1} \pmatrix{1 & 0 \cr u_{1} & 1}
= \pmatrix{1 & 0 \cr u_{1} + u_{2} & 1} .
\end{equation}
We note that the multiplication is commutative, and the parameter
becomes additive.  These matrices convert multiplication into
addition, as logarithmic functions do.

\end{document}